\documentclass[letterpaper, 11pt, conference, onecolumn]{ieeeconf}

 \IEEEoverridecommandlockouts                              
\usepackage{epsfig}
\usepackage{url}
\usepackage{multicol}
\usepackage{amssymb}
\usepackage[tbtags]{amsmath}
\usepackage{latexsym}
\usepackage{euscript}
\usepackage{graphics}

\usepackage{color}

\usepackage[tbtags]{amsmath} 
\usepackage{amssymb}  
\usepackage{verbatim} 

\DeclareMathOperator*{\argmax}{arg\,max}

\newcounter{actr}
{\begin{list}{(\alph{actr})}{\usecounter{actr}}}{\end{list}}

\newcounter{ictr}
{\begin{list}{(\roman{ictr})}{\usecounter{ictr}}}{\end{list}}

\newtheorem{theorem}{Theorem}
 \newtheorem{lemma}{Lemma}
 \newtheorem{definition}{Definition}

\newcommand{\qed}{\rule[0.1ex]{1.4ex}{1.6ex}}

\newcounter{psctr}
\newcounter{probctr}[psctr]

\DeclareMathAlphabet{\mathbsf}{OT1}{cmss}{bx}{n}
\DeclareMathAlphabet{\mathssf}{OT1}{cmss}{m}{sl}

\DeclareSymbolFont{bsfletters}{OT1}{cmss}{bx}{n}
\DeclareSymbolFont{ssfletters}{OT1}{cmss}{m}{n}
\DeclareMathSymbol{\bsfGamma}{0}{bsfletters}{'000}
\DeclareMathSymbol{\ssfGamma}{0}{ssfletters}{'000}
\DeclareMathSymbol{\bsfDelta}{0}{bsfletters}{'001}
\DeclareMathSymbol{\ssfDelta}{0}{ssfletters}{'001}
\DeclareMathSymbol{\bsfTheta}{0}{bsfletters}{'002}
\DeclareMathSymbol{\ssfTheta}{0}{ssfletters}{'002}
\DeclareMathSymbol{\bsfLambda}{0}{bsfletters}{'003}
\DeclareMathSymbol{\ssfLambda}{0}{ssfletters}{'003}
\DeclareMathSymbol{\bsfXi}{0}{bsfletters}{'004}
\DeclareMathSymbol{\ssfXi}{0}{ssfletters}{'004}
\DeclareMathSymbol{\bsfPi}{0}{bsfletters}{'005}
\DeclareMathSymbol{\ssfPi}{0}{ssfletters}{'005}
\DeclareMathSymbol{\bsfSigma}{0}{bsfletters}{'006}
\DeclareMathSymbol{\ssfSigma}{0}{ssfletters}{'006}
\DeclareMathSymbol{\bsfUpsilon}{0}{bsfletters}{'007}
\DeclareMathSymbol{\ssfUpsilon}{0}{ssfletters}{'007}
\DeclareMathSymbol{\bsfPhi}{0}{bsfletters}{'010}
\DeclareMathSymbol{\ssfPhi}{0}{ssfletters}{'010}
\DeclareMathSymbol{\bsfPsi}{0}{bsfletters}{'011}
\DeclareMathSymbol{\ssfPsi}{0}{ssfletters}{'011}
\DeclareMathSymbol{\bsfOmega}{0}{bsfletters}{'012}
\DeclareMathSymbol{\ssfOmega}{0}{ssfletters}{'012}

\newcommand{\svx}{x}            

\newcommand{\svy}{y}

\newcommand{\svz}{z}

\usepackage{fullpage}

\begin{document}

\title{\LARGE \bf  Interference channel capacity region for randomized fixed-composition codes
}

 \author{Cheng Chang   
 \thanks{Cheng Chang  is  with the Hewlett-Packard Laboratories,
 Palo Alto, CA
        {\tt\small Email:  cchang@eecs.berkeley.edu}}%
 }

\maketitle

\begin{abstract}
   The randomized fixe-composition with optimal decoding
   error exponents  are studied~\cite{Raul_ISIT,Raul_journal} for the finite
   alphabet interference channel (IFC) with two transmitter-receiver pairs.
In this paper we  investigate  the capacity region of the randomized
fixed-composition coding scheme.  A complete characterization of the
capacity region of the said coding scheme is given. The inner bound
is derived by showing the existence of a positive error exponent
within the capacity region. A simple universal decoding rule is
given.
 The tight outer bound is  derived by extending a technique first developed in~\cite{Dueck_RC}
for single input output channels to interference channels.
 It is shown that even with a sophisticated time-sharing scheme
 among randomized fixed-composition codes,
  the capacity region of the randomized fixed-composition coding
  is not bigger than the
  known Han-Kobayashi~\cite{Han_Kobayashi} capacity region.
  This suggests that
   the average behavior of random codes are not sufficient to get new capacity regions.
   \\

\end{abstract}

\maketitle

\section{Introduction}
\begin{figure*}[htpb]
\setlength{\unitlength}{3247sp}%
\begingroup\makeatletter\ifx\SetFigFont\undefined%
\gdef\SetFigFont#1#2#3#4#5{%
  \reset@font\fontsize{#1}{#2pt}%
  \fontfamily{#3}\fontseries{#4}\fontshape{#5}%
  \selectfont}%
\fi\endgroup%
\begin{picture}(8255,3049)(-400,-4098)
\thinlines {\put(1501,-1936){\vector( 1, 0){375}}
}%
{\put(3751,-3361){\vector( 1, 0){1275}}
}%
{\put(1501,-3286){\line( 0, 1){  0}} \put(1501,-3286){\vector( 1,
0){375}}
}%
{\put(3751,-2011){\vector( 1,-1){1200}}
}%
{\put(3751,-1936){\vector( 1, 0){1275}}
}%
{\put(1876,-2311){\framebox(975,750){}}
}%
{\put(1876,-3586){\framebox(975,750){}}
}%
{\put(3751,-3286){\vector( 1, 1){1200}}
}%
{\put(2890,-3286){\vector( 1, 0){150}}
}%
{\put(2890,-1936){\vector( 1, 0){150}}
}%

{\put(5026,-2311){\framebox(1425,750){}}
}%
{\put(5026,-3586){\framebox(1425,750){}}
}%
{\put(6451,-1936){\vector( 1, 0){375}}
}%
{\put(6451,-3211){\vector( 1, 0){375}}
}%
{\put(6826,-3586){\framebox(975,750){}}
}%
{\put(6826,-2311){\framebox(975,750){}}
}%
{\put(7801,-1936){\vector( 1, 0){375}}
}%
{\put(7801,-3211){\vector( 1, 0){375}}
}%
\put(5201,-2011){\makebox(0,0)[lb]{\smash{{\SetFigFont{9}{14.4}{\rmdefault}{\mddefault}{\updefault}{$W_{Z|XY}(z|x,y)$ }%
}}}}
\put(3076,-1986){\makebox(0,0)[lb]{\smash{{\SetFigFont{9}{14.4}{\rmdefault}{\mddefault}{\updefault}{$x^n (m_x)$}%
}}}}
\put(3076,-3286){\makebox(0,0)[lb]{\smash{{\SetFigFont{9}{14.4}{\rmdefault}{\mddefault}{\updefault}{$y^n (m_y)$}%
}}}}
\put(-200,-1936){\makebox(0,0)[lb]{\smash{{\SetFigFont{9}{14.4}{\rmdefault}{\mddefault}{\updefault}{$m_x \in \{1,2,...2^{nR_x}\}$}%
}}}}
\put(5201,-3286){\makebox(0,0)[lb]{\smash{{\SetFigFont{9}{14.4}{\rmdefault}{\mddefault}{\updefault}{$\tilde W_{\tilde Z|XY}(\tilde z|x,y)$ }%
}}}}
\put(1951,-3286){\makebox(0,0)[lb]{\smash{{\SetFigFont{9}{14.4}{\rmdefault}{\mddefault}{\updefault}{Encoder Y}%
}}}}
\put(1951,-1936){\makebox(0,0)[lb]{\smash{{\SetFigFont{9}{14.4}{\rmdefault}{\mddefault}{\updefault}{Encoder X}%
}}}}
\put(6901,-2011){\makebox(0,0)[lb]{\smash{{\SetFigFont{9}{14.4}{\rmdefault}{\mddefault}{\updefault}{Decoder X}%
}}}}
\put(6901,-3286){\makebox(0,0)[lb]{\smash{{\SetFigFont{9}{14.4}{\rmdefault}{\mddefault}{\updefault}{Decoder Y}%
}}}}
\put(-200,-3286){\makebox(0,0)[lb]{\smash{{\SetFigFont{9}{14.4}{\rmdefault}{\mddefault}{\updefault}{$m_y \in \{1,2,...2^{nR_y}\}$}%
}}}}
\put(8251,-1936){\makebox(0,0)[lb]{\smash{{\SetFigFont{9}{14.4}{\rmdefault}{\mddefault}{\updefault}{$\widehat{m}_x(z^n)$}%
}}}}
\put(8251,-3286){\makebox(0,0)[lb]{\smash{{\SetFigFont{9}{14.4}{\rmdefault}{\mddefault}{\updefault}{$\widehat{m}_y(\tilde z^n)$}%
}}}}
\put(6551,-1836){\makebox(0,0)[lb]{\smash{{\SetFigFont{9}{14.4}{\rmdefault}{\mddefault}{\updefault}{$z^n$}%
}}}}
\put(6551,-3086){\makebox(0,0)[lb]{\smash{{\SetFigFont{9}{14.4}{\rmdefault}{\mddefault}{\updefault}{$ \tilde z^n $}%
}}}}

\end{picture}%

\caption[ ]{A discrete memoryless interference channel of two users}
    \label{fig.interference_channel}
\end{figure*}
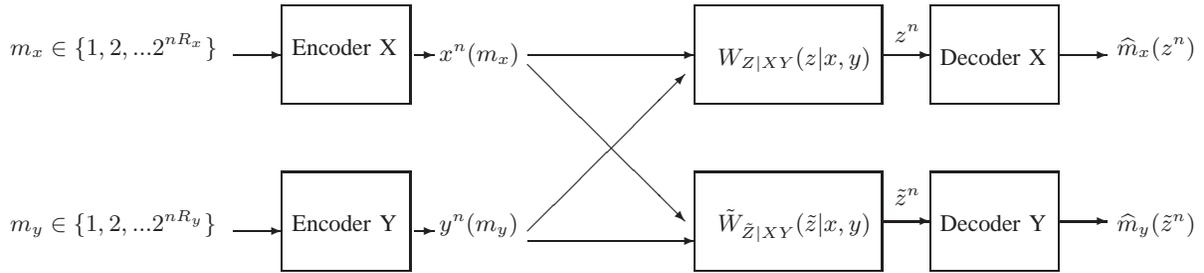

In~\cite{Han_Kobayashi}, the capacity region of interference channel
is studied for both discrete and Gaussian cases. In this paper we
study the discrete interference channels $W_{Z|X,Y}$ and $\tilde
W_{\tilde Z|X,Y}$ with two pairs of encoders and decoders as shown
in Figure~\ref{fig.interference_channel}. The two channel inputs are
$\svx^n\in {\cal X}^n$ and $\svy^n\in {\cal Y}^n$,  outputs are
$\svz^n\in \mathcal Z^n$ and $\tilde z^n\in \tilde{\cal Z}^n$
respectively,  where $\cal X$, $\cal Y$, $\cal Z$ and $\tilde{\cal
Z}$ are finite sets. We study the basic interference channel  where
each encoder only has a private message to the correspondent
decoder.

Some recent progress on the capacity region for Gaussian
interference channels is reported in~\cite{Raul1bit}, however, the
capacity regions for general interference channels are unknown. We
focus our investigation on the capacity region for a specific coding
scheme: randomized fixed-composition codes while the error
probability is defined as the average error over all code book with
a certain composition (type). Fixed-composition coding is a useful
coding scheme in the investigation of both
upper~\cite{Gallager_sphere} and lower bounds of channel coding
error exponents~\cite{Csiszar_graph} for point to point channel and
~\cite{Pokorny_MAC,Liu_huges} for multiple access (MAC) channels.
Recently in~\cite{Raul_ISIT} and~\cite{Raul_journal}, randomized
fixed-composition codes are used to derive a lower bound on the
error exponent for discrete interference channels. A lower bound on
the maximum-likelihood decoding error exponent is derived, this is
 a new attempt in investigating the error exponents for
interference channels. The unanswered question is the capacity
region of such coding schemes.

In this paper, we give a complete characterization of the
interference channel capacity region for randomized
fixed-composition codes.  To prove the achievability of the capacity
region, we prove the positivity everywhere in the capacity region of
a universal decoding error exponent. This error exponent is derived
by the method of types~\cite{Csiszar:98}, in particular the
universal decoding scheme used for multiple-access
channels~\cite{Pokorny_MAC}. A better error exponent can be achieved
by using the more complicated universal decoding rules developed
in~\cite{Liu_huges}. But since they both have the same achievable
capacity region, we use the simpler scheme in~\cite{Pokorny_MAC}. To
prove the the converse, that the achievable region matches the outer
bound, we extend the technique in~\cite{Dueck_RC} for point to point
channels to interference channels by using the known capacity region
results for multiple-access channels. The result reveals the
intimate relations between interference channels and multiple-access
channels. With the capacity region for fixed-composition code
established, it is evident that this capacity region is a subset of
the Han-Kobayashi region~\cite{Han_Kobayashi}.

The technical proof of this paper is focused on the average behavior
of fixed-composition code books. However this fundamental setup can
be generalized in the following three directions.
\begin{itemize}
\item
  It is obvious that there exists a code book that its decoding error
is no bigger than the average decoding error over all code books.
Hence the achievability results in this paper guarantees the
existence of a of deterministic coding scheme with at least the same
error exponents and capacity region. More discussions are in
Section~\ref{sec.average_random}.

\item The focus of this paper is on the fixed-composition codes
 with a composition $P$, where $P$ is a distribution on the input alphabet.
 This code book
generation is different from the non-fixed-composition random
coding~\cite{Gallager} according to distribution $P$. It is well
known in the literature that the  fixed-composition code gives
better error exponent result in low rate regime for point to point
channels~\cite{Csiszar_graph} and multiple-access
channels~\cite{Pokorny_MAC,Liu_huges}. It is the same case for
interference channels and hence the capacity region result in this
paper applies to the non-fixed-composition random codes.

\item  Time-sharing is a key element in achieving capacity regions for
multi-terminal channels~\cite{Cover}. For instance, for
multiple-access channels, simple time-sharing among operational rate
pairs gives the entire capacity
 region.  We show that the our fixed composition codes can be
 used to build a time-sharing
capacity region for interference channel. More interestingly, we
show that the simple time-sharing technique that gives the entire
capacity region for multiple-access channels  is not enough to get
the largest capacity region, a more sophisticated time-sharing
scheme is needed. Detailed discussions are in
Section~\ref{sec.timeshare}.

\end{itemize}

The outline of the paper is as follows. In
Section~\ref{sec.setupmainresult} we first formally define
randomized  fixed-composition codes and its capacity region and then
in Section~\ref{sec.mainresult} we present the main result of this
paper: the   interference channel capacity region  for randomized
fixed-composition code in Theorem~\ref{Thm.Int_capacity}. The proof
is later shown in Section~\ref{sec.proof} with more details in the
appendix.  Finally in Section~\ref{sec.timeshare}, we argue that due
to the non-convexity of the randomized fixed-composition coding, a
more sophisticated time-sharing scheme is needed. This shows the
necessity of studying the geometry of the code-books for
interference channels.

\section{Randomized fixed-composition code and its capacity
region}\label{sec.setupmainresult}

We first review the definition of randomized fixed-composition code
that is studied intensively in previous works. Then the definition
of the interference channel capacity region for such codes is
introduced. Then we give the main result of this paper: the complete
characterization of the capacity region for randomized
fixed-composition codes.

\subsection{Randomized fixed-composition codes}
A randomized fixed-composition code is a uniform distribution on the
code books in which every codeword is from the type set with the
fixed composition (type).

First we introduce the notion of type set~\cite{Cover}.  A type set
$\mathcal T^n(P)$ is  a set of all the strings $x^n\in \mathcal X^n$
with the same type $P$ where $P$ is a probability
distribution~\cite{Cover}. A sequence of type sets $\mathcal T^n
\subseteq \mathcal X^n$ has composition $P_X$ if the types of
$\mathcal T^n$ converges to $P_X$, i.e. $\lim\limits_{n\rightarrow
\infty} \frac{N(a|\mathcal T^n)}{n}= P_X(a)$ for all $a\in \cal X$
that $P_X(a)>0$ and $N(a|\mathcal T^n)=0$ for all $a\in \cal X$ that
$P_X(a)=0$, where $N(a|\mathcal T^n)$ is the number of occurrence of
$a$ in type $\mathcal T^n$.  We ignore the nuisance of the integer
effect  and assume that $n P_X(a)$ is an integer for all $a\in \cal
X$ and $nR_x$ and $nR_y$ are also integers. This is indeed a
reasonable assumption since we study long block length $n$ and all
the information theoretic quantities studied in this paper are
continuous on the code compositions and rates. We  simply denote by
$\mathcal T^n(P_X)$ the length-$n$ type set which has ``asymptotic''
type $P_X$, later in the appendix we abuse the notations by simply
writing $x^n\in P_X$ instead of $x^n\in \mathcal T^n(P_X)$.
Obviously, there are $|\mathcal T^n(P_X)|^{2^{nR_x}}$ many code
books with fixed-composition $P_X$ and rate $R_x$

In this paper, we study the  randomized fixed-composition codes,
where each code book with all codewords from the fixed composition
being chosen with the same probability. Equivalently,  over all
these code books, a code word for message $i$ is uniformly i.i.d
distributed on the type set $\mathcal T^n(P_X)$. A formal definition
is as follows. \vspace{0.1in}

\begin{definition}{Randomized fixed-composition codes}\label{def:randomized_coding}:
 for a probability distribution  $P_X$ on $\cal X$, a rate $R_x$
 randomized fixed-composition-$P_X$ encoder picks a code book with the
 following probability, for any fixed-composition-$P_X$ code book
  $\theta^n =(\theta^n(1),\theta^n(2),..., \theta(2^{nR_x}))$,  where
   $\theta^n(i)\in \mathcal T^n(P_X)$,  $i=1,2,..., 2^{nR_x}$, and $\theta^n(i)$
   and $\theta^n(j)$ may not be different for $i\neq j$,
   the code book $\theta_n$ is chosen, i.e. $x^n(i)=\theta^n(i), \ \ i=1,2,...,2^{nR_x}$, with probability
\begin{eqnarray*}
\left (\frac{1}{|\mathcal T^n(P_X)|}\right)^{2^{nR_x}}
\end{eqnarray*}
In other words, the choice of the code book is a random  variable
$c_X$ uniformly distributed on the index set of all the possible
code books with fixed-composition $P_X$: $\{1,2,3,...,|\mathcal
T^n(P_X)|^{2^{nR_x}}\}$, while $c_X$ is shared between the encoder
$X$ and the decoders $X$ and $Y$. \vspace{0.1in}
\end{definition}

The key property  of the randomized fixed-composition code is that
for any message subset $\{i_1, i_2,...i_l\}\subseteq \{1,2,...,
2^{nR_x}\}$, the code words for these messages are identical
independently distributed on the type set of $\mathcal T^n(P_X)$.

For randomized fixed-composition codes, the average error
probability $P_{e(x)}^n(R_x, R_y, P_X, P_Y)$ for X is the
expectation of decoding error over all message, code books and
channel behaviors.
\begin{eqnarray}
P_{e(x)}^n(R_x, R_y, P_X, P_Y) &&= \left (\frac{1}{|\mathcal
T^n(P_X)|}\right)^{2^{nR_x}}
\left (\frac{1}{|\mathcal T^n(P_Y)|}\right)^{2^{nR_y}}\label{eqn.inter_error_avg_X}\\
\sum_{c_X}\sum_{c_Y}\frac{1}{2^{nR_x}}&&\sum_{m_x}\frac{1}{2^{nR_y}}\sum_{m_y}\sum_{z^n}
 W_{Z|XY}(z^n|x^n(m_x),y^n(m_y))
1(\widehat m_x(z^n)\neq m_x)\nonumber
\end{eqnarray}

where $x^n(m_x)$ is the code word of message $m_x$ in code book
$c_X$, similarly for $y^n(m_y)$, $\widehat m_x(z^n)$ is the decision
made by the decoder knowing the code books $c_X$ and $c_Y$.

\subsection{Randomized fixed-composition coding capacity for interference channels}

Given the definitions of randomized fixed-composition coding  and
the average error probability in (\ref{eqn.inter_error_avg_X}) for
such codes, we can formally define the capacity region for such
codes. \vspace{0.1in}
\begin{definition}{Capacity region for randomized fixed-composition codes}: for
 a fixed-composition $P_X$ and $P_Y$, a rate pair $(R_x,R_y)$ is said to be achievable for $X$,
 if for all $\delta>0$, there exists $N_\delta<\infty$,  s.t. for all $n>N_\delta$,
\begin{eqnarray}
P_{e(x)}^n(R_x, R_y, P_X, P_Y)< \delta \mbox{}
\end{eqnarray}
We denote by $\mathcal R_x(P_X,P_Y)$ the closure of the union of the
all achievable rate pairs. Similarly we  denote by $\mathcal
R_y(P_X,P_Y)$ the achievable region for $Y$,  and $\mathcal
R_{xy}(P_X,P_Y)$ for  $(X,Y)$ where both decoding errors are small.
Obviously
\begin{eqnarray}
\mathcal R_{xy}(P_X,P_Y)= \mathcal R_{x}(P_X,P_Y)\bigcap\mathcal
R_{y}(P_X,P_Y) .\label{eqn.union_xy}
\end{eqnarray}
\end{definition}
\vspace{0.1in} We only need to focus our investigation on $\mathcal
R_{x}(P_X,P_Y)$, then by the obvious symmetry, both $\mathcal
R_{y}(P_X,P_Y)$ and $\mathcal R_{xy}(P_X,P_Y)$  follow.

\subsection{Capacity region of the fixed-composition code, $\mathcal
R_{x}(P_X,P_Y)$, for $X$ }\label{sec.mainresult}

The main result of this paper is the complete characterization of
the randomized fixed-composition capacity region  $\mathcal
R_{x}(P_X,P_Y)$ for $X$, as illustrated in~(\ref{eqn.union_xy}), by
symmetry, $\mathcal R_{xy}(P_X,P_Y)$ follows.

\begin{figure*}
\setlength{\unitlength}{3247sp}%
\begingroup\makeatletter\ifx\SetFigFont\undefined%
\gdef\SetFigFont#1#2#3#4#5{%
  \reset@font\fontsize{#1}{#2pt}%
  \fontfamily{#3}\fontseries{#4}\fontshape{#5}%
  \selectfont}%
\fi\endgroup%
\begin{picture}(5562,4416)(-1701,-4240)
\thinlines { \put(1951,-3961){\vector( 1, 0){4800}}
}%
{ \put(1951,-3961){\vector( 0, 1){3825}}
}%
\thicklines { \put(3601,-211){\line( 0,-1){1350}}
\put(3601,-1561){\line( 1,-1){1500}} \put(5101,-3061){\line(
0,-1){900}}
}%
{
\multiput(5101,-3061)(0.00000,89.9054){32}{\makebox(6.6667,10.0000){\SetFigFont{10}{12}{\rmdefault}{\mddefault}{\updefault}.}}
}%
{
\multiput(1951,-1561)(90.00000,0.00000){35}{\makebox(6.6667,10.0000){\SetFigFont{10}{12}{\rmdefault}{\mddefault}{\updefault}.}}
}%
\put(1626,-361){\makebox(0,0)[lb]{\smash{{\SetFigFont{9}{14.4}{\rmdefault}{\mddefault}{\updefault}{ $R_y$}%
}}}}
\put(6301,-3836){\makebox(0,0)[lb]{\smash{{\SetFigFont{9}{14.4}{\rmdefault}{\mddefault}{\updefault}{ $R_x$}%
}}}}
\put(2701,-811){\makebox(0,0)[lb]{\smash{{\SetFigFont{9}{14.4}{\rmdefault}{\mddefault}{\updefault}{ $I$}%
}}}}
\put(2701,-2761){\makebox(0,0)[lb]{\smash{{\SetFigFont{9}{14.4}{\rmdefault}{\mddefault}{\updefault}{ $II$}%
}}}}
\put(4576,-811){\makebox(0,0)[lb]{\smash{{\SetFigFont{9}{14.4}{\rmdefault}{\mddefault}{\updefault}{ $III$}%
}}}}
\put(4576,-1861){\makebox(0,0)[lb]{\smash{{\SetFigFont{9}{14.4}{\rmdefault}{\mddefault}{\updefault}{ $IV$}%
}}}}
\put(6001,-1861){\makebox(0,0)[lb]{\smash{{\SetFigFont{9}{14.4}{\rmdefault}{\mddefault}{\updefault}{ $V$}%
}}}}
\put(3301,-4186){\makebox(0,0)[lb]{\smash{{\SetFigFont{9}{14.4}{\rmdefault}{\mddefault}{\updefault}{ $I(X;Z)$}%
}}}}
\put(1076,-3111){\makebox(0,0)[lb]{\smash{{\SetFigFont{9}{14.4}{\rmdefault}{\mddefault}{\updefault}{ $I(Y;Z)$}%
}}}}
\put(901,-1661){\makebox(0,0)[lb]{\smash{{\SetFigFont{9}{14.4}{\rmdefault}{\mddefault}{\updefault}{ $I(Y;Z|X)$}%
}}}}
\put(4626,-4186){\makebox(0,0)[lb]{\smash{{\SetFigFont{9}{14.4}{\rmdefault}{\mddefault}{\updefault}{ $I(X;Z|Y)$}%
}}}}
\end{picture}%

\caption[ ]{Randomized fixed-composition capacity region $\mathcal
R_x(P_X, P_Y)$ for $X$, the achievable region is the union of Region
$I$ and $II$.}
    \label{fig.inter_region}
\end{figure*}
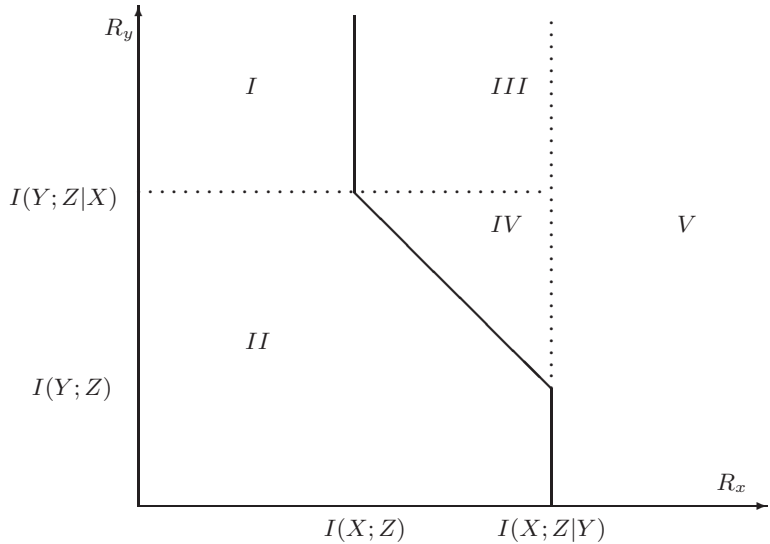
\vspace{0.1in}

\begin{theorem}{Interference channel capacity region $\mathcal R_{x}(P_X,P_Y)$ for
 randomized fixed-composition codes with compositions $P_X$ and $P_Y$:}
\label{Thm.Int_capacity}
\begin{eqnarray}\label{eqn.int_region}
\mathcal R_x(P_X, P_Y)&=&\{(R_x,R_y): 0\leq R_x< I(X;Z), 0\leq R_y\}\ \ \ \bigcup\nonumber\\
&& \{(R_x,R_y): 0\leq R_x< I(X;Z|Y),   R_x+R_y< I(X,Y;Z)\}
\end{eqnarray}
where the random variables in (\ref{eqn.int_region}), $(X, Y, Z)\sim
P_X P_Y W_{Z|X,Y}$.  The region $\mathcal R_{x}(P_X,P_Y)$ is
illustrated in Figure~\ref{fig.inter_region}.
\end{theorem}
\vspace{0.1in}

 The achievable part of the theorem states
that:  for a rate pair $(R_x, R_y)\in \mathcal R_x(P_X, P_Y)$, the
union of  Region $I$ and $II$ in Figure~\ref{fig.inter_region}, for
all $\delta
>0$, there exists $N_\delta<\infty$, s.t. for all $n>N_\delta$, the
average error probability (\ref{eqn.inter_error_avg_X}) for the
randomized code from compositions $P_X $ and $P_Y$ is smaller than
$\delta$ for $X$:
$$P_{e(x)}^n(R_x, R_y, P_X, P_Y)< \delta $$ for some decoding rule.
Region $II$ is also the multiple-access capacity region for
fixed-composition codes $(P_X, P_Y)$ for channel $W_{Z|XY}$.

\vspace{0.1in}

The converse of the theorem states that for any rate pair $(R_x,
R_y)$  outside of  $\mathcal R_x(P_X, P_Y)$, that is region $III$,
$IV$ and $IV$ in Figure~\ref{fig.inter_region}, there exists
$\delta>0$, such that for all $n$,
$$P_{e(x)}^n(R_x, R_y, P_X, P_Y)> \delta $$ no matter what decoding rule is
used. Note that the definition of the error probability
$P_{e(x)}^n(R_x, R_y, P_X, P_Y)$ defined
in~(\ref{eqn.inter_error_avg_X})

The proof of Theorem~\ref{Thm.Int_capacity} is  in
Section~\ref{sec.proof}.

\begin{figure}
\begin{center}
\includegraphics[width=90mm]{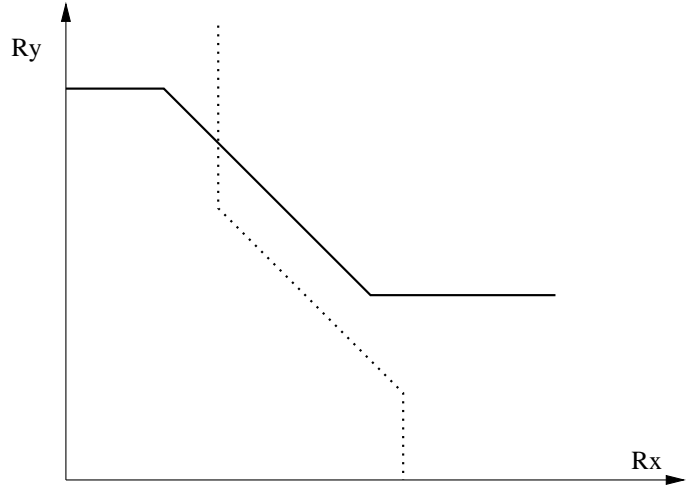}
\end{center}
\caption[ ]{A typical randomized fixed-composition capacity region
$\mathcal R_{xy}(P_X, P_Y)= \mathcal R_x(P_X, P_Y) \cap \mathcal
R_y(P_X, P_Y)$ is the intersection of the dotted line and the solid
lines, this capacity region is not necessarily convex. }
    \label{fig.inter_region_XY}
\end{figure}

\subsection{Necessities of  more sophisticated  time-sharing
schemes}\label{sec.time-sharing-firstrun}

In the achievability part of Theorem~\ref{Thm.Int_capacity}, we
prove that the average error probability for $X$ is arbitrarily
small for a randomized fixed-composition code if the rate pair
$(R_x,R_y)$ is inside the capacity region $\mathcal R_x(P_X, P_Y)$.
For interference channels, it is obvious that the rate region for
both $X$  and $Y$ is:
\begin{eqnarray}\mathcal R_{xy}(P_X, P_Y)=
\mathcal R_x(P_X, P_Y) \cap \mathcal R_y(P_X,
P_Y),\label{eqn.XYREGION}
\end{eqnarray}

where $\mathcal R_y(P_X, P_Y)$ is defined in the same manner as
$\mathcal R_x(P_X, P_Y)$ but the channel is $ \tilde W_{\tilde
Z|XY}$ instead of $W_{Z|XY}$ as shown in
Figure~\ref{fig.interference_channel}. A typical capacity region
$\mathcal R_{xy}(P_X, P_Y)$ is shown in
Figure~\ref{fig.inter_region_XY}. It is not necessarily convex.

However, by  a simple time-sharing between different rate pairs for
the same composition, we can convexify the capacity region. Then the
convex hull of the union of all such capacity regions of different
compositions gives    a bigger convex achievable  capacity region.
This capacity region of the interference channel is
\begin{eqnarray}
CONVEX\left(\bigcup_{P_X, P_Y} \mathcal
R_{xy}(P_X,P_Y)\right).\nonumber
\end{eqnarray}

It is tempting to claim that the above convex capacity region is the
largest one can get by time-sharing  the ``basic'' fixed-composition
codes as multiple-access channels shown in~\cite{Cover}. However, as
will be discussed later in Section~\ref{sec.timeshare}, it is not
the case. A more sophisticated time-sharing gives a bigger capacity
region.

This is an important difference between interference channel coding
and multiple-access channel coding because the fixed-composition
capacity region is convex for the latter and hence the simple
time-sharing gives the biggest capacity region~\cite{Cover}.
Time-sharing capacity is detailed in Section~\ref{sec.timeshare}.

\subsection{Existence of a good code for an interference
channel}\label{sec.average_random}

In this paper we focus our study on the average (over all messages)
error probability over all code books with the same composition. For
a rate pair $(R_x,R_y)$, if the average error probability for $X$ is
smaller than $\delta$, then obviously there exists a code book such
that the error probability is smaller than $\delta$ for $X$. This
should be clear from the definition of error probability
$P_{e(x)}^n(R_x, R_y, P_X, P_Y)$ in~(\ref{eqn.inter_error_avg_X}).
In the following example, we illustrate that this is also the case
for decoding error for both $X$ and $Y$. We claim without proof that
this is also true for ``uniform'' time-sharing coding schemes later
discussed in Section~\ref{sec.timeshare}. The existence of a code
book that achieves the error exponents in the achievability
 part of the proof of Theorem~\ref{Thm.Int_capacity} can
also be shown. The proof is similar to that in~\cite{Gallager} and
Exercise 30~(b) on page 198~\cite{Csiszar}.

  Similar to the
error probability for $X$ defined in~(\ref{eqn.inter_error_avg_X}),
we define the average joint error probability for  $X$ and $Y$ as
\begin{eqnarray}
  P_{e(xy)}^n(R_x, R_y, P_X, P_Y)  &=& \left (\frac{1}{|\mathcal
T^n(P_X)|}\right)^{2^{nR_x}}\left (\frac{1}{|\mathcal
T^n(P_Y)|}\right)^{2^{nR_y}}
 \sum_{c_X}\sum_{c_Y}\frac{1}{2^{nR_x}}\sum_{m_x}\frac{1}{2^{nR_y}}\sum_{m_y} \label{eqn.inter_error_avg_XY}\\
&&\ \ \ \big\{ \sum_{z^n} W_{Z|XY}(z^n|x^n(m_x),y^n(m_y)) 1(\widehat
m_x(z^n)\neq m_x)
\nonumber\\
&& \ \ \ \ \ \  + \sum_{\tilde z^n} \tilde W_{\tilde Z|XY}(\tilde
z^n|x^n(m_x),y^n(m_y)) 1(\widehat m_y(\tilde  z^n)\neq
m_y)\big\}\nonumber
\end{eqnarray}

For a rate pair $(R_x,R_y)\in \mathcal R_{xy}(P_X, P_Y)= \mathcal
R_x(P_X, P_Y)\bigcap \mathcal R_y(P_X, P_Y) $. We know that for all
$\delta
>0$, there exists $N_\delta<\infty$, s.t. for all $n>N_\delta$, the
average error probability   is smaller than $\delta$ for user $X$
and user $Y$:\\
 $P_{e(x)}^n(R_x, R_y, P_X, P_Y)< \delta $ and
  $P_{e(y)}^n(R_x, R_y, P_X, P_Y)< \delta
 $. It is easy to see that the average joint error probability for
 user
 $X$ and $Y$ can be bounded by:
\begin{eqnarray}
P_{e(xy)}^n(R_x, R_y, P_X, P_Y) &=& P_{e(x)}^n(R_x, R_y, P_X, P_Y)+
P_{e(y)}^n(R_x, R_y, P_X,
P_Y)\nonumber\\
&\leq&  2\delta\label{eqn.union_bound_onxy}
\end{eqnarray}
From (\ref{eqn.inter_error_avg_XY}), we know that $P_{e(xy)}^n(R_x,
R_y, P_X, P_Y)$ is the average error probability of \textit{all}
$(P_X, P_Y)$-fixed-composition codes. Together with
(\ref{eqn.union_bound_onxy}), we know that there exists at least
\textit{one} code book such that the error probability is no bigger
than $2 \delta$.


Note, the converse of the randomized coding does not guarantee that
there is not a single good fixed-composition code book.  The
converse claims that, the average (over all code books with the
composition) decoding error probability does not converge to zero if
the rate pair is outside the capacity region in
Theorem~\ref{Thm.Int_capacity}.

\section{Proof of Theorem~\ref{Thm.Int_capacity}}\label{sec.proof}

There are two parts of the theorem,  achievability  and converse.
The achievability part is proved by applying the classical method of
types  in point to point channel coding and MAC channel coding for
randomized fixed-composition code. The converse is proved by
extending the technique first developed in~\cite{Dueck_RC} for point
to point channels to interference channels.

\subsection{Achievability}
We show that in the interior of the capacity region, i.e. the union
of Region $I$ and $II$ in Figure~\ref{fig.inter_region}, a positive
error exponent is achieved by applying the randomized
fixed-composition coding defined in
Definition~\ref{def:randomized_coding}.  In
Sections~\ref{section:regionII}  and~\ref{section:regionI}, we
describe the universal decoding rules for Region $II$ and $I$
respectively. We then present the error exponent results in
Lemma~\ref{lemma:regionII} in Section~\ref{section:regionIIEE} and
Lemma~\ref{lemma:regionI} in Section~\ref{section:regionIEE} that
covers Region $II$ and $I$ respectively. Then in
Lemma~\ref{lemma:positiveness} in Section~\ref{sec.positivityEE}, we
show that these error exponents are positive in the interior of the
capacity region $\mathcal R_x(P_X, P_Y)$ and hence conclude the
proof of the achievability part in Theorem~\ref{Thm.Int_capacity}.
 \vspace{0.1in}

\subsubsection{Decoding rule in Region $II$} \label{section:regionII} In Region $II$, we show that
decoder $X$ can decode  both message $m_x$ and $m_y$ with small
error probabilities. This is essentially a multiple-access channel
coding problem.   We use the technique developed in~\cite{Csiszar}
 to derive the positive error exponents  that parallel to those in~\cite{Pokorny_MAC}.
 The decoder is a simple maximum mutual
information\footnote{A more sophisticated decoding rule based on
minimum conditional entropy decoding for multiple-access channel is
developed in~\cite{Liu_huges}, it is shown that this decoding rule
achieves a bigger error exponent in low rate regime. The goal of
this paper is, however, not to derive the tightest lower bound on
the error exponent. We only need a coding scheme to achieve positive
error exponent in the capacity region in
Theorem~\ref{Thm.Int_capacity}. Hence we use the simpler decoding
rule here. } decoder~\cite{Csiszar}. This decoding rule is universal
in the sense that the decoder does not need to know the multiple
access channel $W_{Z|XY}$. We describe the decoding rule here, the
estimate of the joint message is the message pair such that the
input to the channel $W_{Z|XY}$ and the output of the channel have
the maximal empirical mutual information. i.e.:
\begin{eqnarray}
(\widehat m_x(z^n), \widehat m_y(z^n))=\argmax_{i\in\{1,2,...,
2^{nR_x}\}, j\in\{1,2,..., 2^{nR_y}\}} I(z^n; x^n(i),
y^n(j))\label{eqn.decoder1}
\end{eqnarray}
where $z^n$ is the channel output and $x^n(i)$ and $y^n(j)$ are the
channel inputs for message $i$ and $j$ respectively. $I(z^n; x^n,
y^n )$ is the empirical mutual information between $z^n$ and $(x^n,
y^n)$, the point to point maximal mutual mutual information decoding
is studied in~\cite{Csiszar}.

 If there is a tie, the decoder can choose an arbitrary winner
or simply declare error. In Lemma~\ref{lemma:regionII}, we show that
 by using the randomized fixed-composition encoding and the maximal
mutual information decoding, a non-negative error exponent is
achieved in Region $II$.

\vspace{0.1in}

\subsubsection{Decoding rule in Region $I$}  \label{section:regionI}In Region $I$,  decoder
$X$ only  estimates $m_x$   by treating the input of encoder $Y$ as
a source of random noises. This is essentially a point to point
channel coding problem. The channel itself has memory since the
input of encoder $Y$ is not memoryless. Similar to the multiple
access channel coding problem studied in Region $II$, we use a
maximal mutual information decoding rule:
\begin{eqnarray}
\widehat m_x (z^n)=\argmax_{i\in\{1,2,..., 2^{nR_x}\} } I(z^n;
x^n(i))\label{eqn.decoder2}
\end{eqnarray}
 In Lemma~\ref{lemma:regionI}, we show that
  by using the randomized fixed-composition encoding and the
maximal mutual information decoding, a non-negative error exponent
is achieved in Region $I$.

\vspace{0.1in}

\subsubsection{Lower bound on the error exponent in Region $II$}\label{section:regionIIEE}
\begin{lemma}{(Region $II$) Multiple-access channel error exponents (joint error probability).}\label{lemma:regionII}
 For the randomized coding scheme described in
Definition~\ref{def:randomized_coding}, and the decoding rule
described in (\ref{eqn.decoder1}),
 the decoding error probability averaged over all messages, code books and channel behaviors is
 upper bounded by an exponential term:
\begin{eqnarray}
&&\Pr((m_x,m_y)\neq (\widehat m_x, \widehat m_y ))\nonumber\\
&=&\left (\frac{1}{|\mathcal T^n(P_X)|}\right)^{2^{nR_x}}\left (\frac{1}{|\mathcal T^n(P_Y)|}\right)^{2^{nR_y}}\label{eqn.inter_error_mac} \\
&&\ \
\sum_{c_X}\sum_{c_Y}\frac{1}{2^{nR_x}}\sum_{m_x}\frac{1}{2^{nR_y}}\sum_{m_y}\sum_{z^n}
 W_{Z|XY}(z^n|x^n(m_x),y^n(m_y))
1\left((\widehat m_x(z^n), \widehat m_y(z^n))\neq (m_x, m_y)\right)\nonumber\\
&\leq& 2^{-n (E-\epsilon_n)}.
\end{eqnarray}

 $\epsilon_n$ converges to zero as $n$ goes to infinity, and $E=\min \{E_{xy}, E_{x|y}, E_{y|x}\},\mbox{
 where}$

\begin{eqnarray} E_{xy}&=&\min_{Q_{XYZ}:Q_X=P_X,
Q_Y=P_Y} D(Q_{Z|XY}\|W| Q_{XY})
+D(Q_{XY}\|P_X\times P_Y)  +|I_Q(X,Y;Z)-R_x-R_y|^+\nonumber\\
E_{x|y}&=&\min_{Q_{XYZ}:Q_X=P_X, Q_Y=P_Y} D(Q_{Z|XY}\|W| Q_{XY})
 + D(Q_{XY}\|P_X\times P_Y)   +|I_Q(X;Z|Y)-R_x|^+\nonumber\\
E_{y|x}&=&\min_{Q_{XYZ}:Q_X=P_X, Q_Y=P_Y} D(Q_{Z|XY}\|W| Q_{XY})  +
D(Q_{XY}\|P_X\times P_Y)
   +|I_Q(Y;Z|X)-R_y|^+\nonumber
\end{eqnarray}
where $|t|^+ =\max\{0,t \}$ and the random variables $(X,Y,Z)\sim
Q_{XYZ}$ in $I_Q(X;Z|Y), I_Q(Y;Z|X)$ and $I_Q(X, Y;Z)$.
\end{lemma}
\vspace{0.1in}

 {\em Remark 1: it is easy to verify that
$D(Q_{Z|XY}\|W| Q_{XY})+D(Q_{XY}\|P_X\times P_Y) = D(Q_{XYZ}\|
P_X\times P_Y\times W)$, so the expressions for the  error exponents
can be further simplified.   We use the expressions similar to those
in~\cite{Pokorny_MAC}  because they are more intuitive. \em}

 {\em Remark 2: The proof parallels that
in~\cite{Pokorny_MAC} which is in turn an extension to the point to
point channel coding problem studied in~\cite{Csiszar}. The method
of types is the main tool for the proofs.  The difference is that we
need to show the lower bound to the average error probability
instead of showing the existence of \textit{a} good code book
in~\cite{Pokorny_MAC}.  Without giving details, we follow Gallager's
proof in~\cite{Gallager} and claim the existence of \textit{a} good
code with the same error exponent as that in~\cite{Pokorny_MAC} as a
simple corollary of Lemma~\ref{lemma:regionII}. \em}

\vspace{0.1in}

\proof First we have an obvious upper bound on the error probability
\begin{eqnarray}
&&\Pr((m_x,m_y)\neq (\widehat m_x, \widehat m_y ))\nonumber\\
&=& \Pr( m_x\neq \widehat m_x, m_y\neq \widehat m_y )  +  \Pr(
m_x\neq \widehat m_x, m_y= \widehat m_y )
+  \Pr( m_x = \widehat m_x, m_y\neq \widehat m_y )\nonumber\\
&\leq&\Pr( m_x\neq \widehat m_x, m_y\neq \widehat m_y ) +  \Pr(
m_x\neq \widehat m_x| m_y= \widehat m_y)
 +  \Pr( m_y\neq \widehat m_y |m_x = \widehat m_x
))\label{eqn:unionbound}
\end{eqnarray}
The inequality~(\ref{eqn:unionbound}) follows the equality
$P(A,B)=P(A|B)P(B)\leq P(A|B)$. Now we upper bound each individual
error probability in (\ref{eqn:unionbound}) respectively by
exponentials of $n$. We only need to show that
\begin{eqnarray}
&&\Pr( m_x\neq \widehat m_x, m_y\neq \widehat m_y )\leq 2^{-n(E_{xy}-\epsilon_n)}, \label{eqn.proofpart1}\\
\mbox{ }&&  \Pr( m_x \neq \widehat m_x| m_y = \widehat m_y )\leq
2^{-n(E_{x|y}-\epsilon_n)},\label{eqn.proofpart2}\\
\mbox{and }&&  \Pr( m_y \neq \widehat m_y| m_x = \widehat m_x )\leq
2^{-n(E_{y|x}-\epsilon_n)}.\label{eqn.proofpart2.aa}
\end{eqnarray}
We  prove (\ref{eqn.proofpart1}) and (\ref{eqn.proofpart2}),
(\ref{eqn.proofpart2.aa}) follows (\ref{eqn.proofpart2}) by
symmetry. The proofs are in Appendix~\ref{section.appendix1}, where
a standard method of type argument is used. \hfill$\square$

\vspace{0.1in}

\subsubsection{Lower bound on the error exponent in Region $I$}\label{section:regionIEE}

\begin{lemma}{(Region $I$) point to point channel coding error exponent (decoding $X$ only).}\label{lemma:regionI} For the randomized coding scheme described in
Definition~\ref{def:randomized_coding}, and the decoding rule
described in (\ref{eqn.decoder2}),
 the decoding error probability averaged over all messages, code books and channel behaviors is
 upper bounded by an exponential term:
\begin{eqnarray}
\Pr(m_x\neq \widehat m_x)&=&\left (\frac{1}{|\mathcal T^n(P_X)|}\right)^{2^{nR_x}}\left (\frac{1}{|\mathcal T^n(P_Y)|}\right)^{2^{nR_y}}\nonumber\\
&&\ \
\sum_{c_X}\sum_{c_Y}\frac{1}{2^{nR_x}}\sum_{m_x}\frac{1}{2^{nR_y}}\sum_{m_y}\sum_{z^n}
  W_{Z|XY}(z^n|x^n(m_x),y^n(m_y))
 1\left(\widehat m_x(z^n)\neq m_x\right)\nonumber\\
&\leq& 2^{-n (E_x-\epsilon_n)}.\label{eqn.inter_error_avg}
\end{eqnarray}

 $\epsilon_n$ converges to zero as $n$ goes to infinity, and
\begin{eqnarray}
E_{x}&=&\min_{Q_{XYZ}:Q_X=P_X, Q_Y=P_Y} D(Q_{Z|XY}\|W| Q_{XY})  +
D(Q_{XY}\|P_X\times P_Y)  +|I_Q(X;Z)-R_x|^+\nonumber
\end{eqnarray}

\end{lemma}
\vspace{0.1in}

\proof We give a unified proof for (\ref{eqn.proofpart1}),
(\ref{eqn.proofpart2}) and (\ref{eqn.inter_error_avg}) in
Appendix~\ref{section.appendix1}. \hfill$\square$ \vspace{0.1in}

With Lemma~\ref{lemma:regionII} and Lemma~\ref{lemma:regionI}, we
know that  some non-negative error exponents can be achieved for the
randomized $(P_X,P_Y)$ fixed-composition code if the rate pair
$(R_x,R_y)\in \mathcal R_x(P_X,P_Y)$. This is because both
Kullback-Leibler divergence and $|\cdot|^+$ are always non-negative.
Now we only need to show the positiveness of those error exponents
when the rate pair is in  the interior of $ \mathcal R_x(P_X,P_Y)$.

\vspace{0.1in}
\subsubsection{Positiveness of the error exponents}\label{sec.positivityEE}
\begin{lemma}{}\label{lemma:positiveness}
For rate pairs $(R_x,R_y)$ in the interior of $ \mathcal
R_x(P_X,P_Y)$ defined in
 Theorem~\ref{Thm.Int_capacity}:
$$\max\{\min\{E_{xy}, E_{x|y}, E_{y|x}\} , E_{x}\} >0.$$
More specifically, we show two things. First, if $R_x< I(X,Z)$,
where $(X,Z)\sim P_{X}\times P_{Y}\times W_{Z|XY}$, then $E_x>0$.
This covers Region $I$.
 Secondly, if $R_x< I(X,Z|Y)$, $R_y< I(Y,Z|X)$ and $R_x+R_y< I(X,Y;Z)$,
 where $(X,Y,Z)\sim P_{X}\times P_{Y}\times W_{Z|XY}$, then $\min\{E_{xy}, E_{x|y},
 E_{y|x}\}>0$,  this covers Region $II$.\\

\proof  First,  suppose that for  some $R_x< I(X,Z)$, $E_x\leq 0$.
Since both Kullback-Leibler divergence and $|\cdot|^+$ are
non-negative functions, we must have $E_x=0$ and hence there exists
a distribution $Q_{XYZ}$, s.t. $Q_X=P_X$, $Q_Y=P_Y$ and all the
individual non-negative functions are zero:
\begin{eqnarray}
  D(Q_{XY}\|P_X\times P_Y)&=&0  \nonumber\\
  D(Q_{Z|XY}\|W| Q_{XY})&=&0\nonumber\\
|I_Q(X;Z)-R_x|^+&=&0\nonumber
\end{eqnarray}
The first equation tells us that $Q_{XY}= P_{X}\times P_Y$. Then the
second equation becomes $  D(Q_{Z|XY}\|W| P_{X}\times P_{Y})=0$,
this means that $Q_{Z|XY}\times  P_{X}\times P_{Y}= W\times
P_{X}\times P_{Y}$, so $I_Q(X;Z)= I(X;Z)$ where the random variables
$(X,Y,Z)\sim  P_{X}\times P_{Y}\times W_{Z|XY} $ in $I(X;Z)$. Now
the third equation becomes $|I(X;Z)-R_x|^+=0$ which is equivalent to
$I(X;Z)\leq  R_x$, this is a contradiction to the fact that $R_x<
I(X,Z)$.

Secondly, suppose that for some rate pair $(R_x, R_y)$ in Region
$II$, i.e. $R_x< I(X,Z|Y)$, $R_y< I(Y,Z|X)$ and $R_x+R_y< I(X,Y;Z)$
and $\min\{E_{xy}, E_{x|y},
 E_{y|x}\}\leq 0$, then $\min\{E_{xy}=0 $ or $ E_{x|y}=0$ or
 $E_{y|x}\}=0$. Following exactly the same argument as that
  in the first part of the proof of Lemma~\ref{lemma:positiveness},
 we can get contradictions with the fact that the rate pair $(R_x,
 R_y)$ is in the interior of Region $II$.\hfill $\square$

\end{lemma}
\vspace{0.1in} From the above three lemmas, we conclude that the
error probability for decoding message $X$ is upper bounded by
$2^{-n(E-\epsilon_n)}$ for all $(R_x,R_y)\in \mathcal R_x(P_X,P_Y)$,
where $E>0$ and $\lim\limits_{n\rightarrow \infty}\epsilon_n=0$.
Hence the error probability converges to zero exponentially fast for
large $n$. This concludes the achievability part of the proof for
Theorem~\ref{Thm.Int_capacity}.

\subsection{Converse}
We show that the average decoding error of Decoder $X$ does not
converge to zero with increasing $n$ if the rate pair $(R_x, R_y)$
is outside the capacity region $\mathcal R_x(P_X, P_Y)$ shown in
Figure~\ref{fig.inter_region}. There are three parts of the proof
for Regions $V$, $IV$ and $III$ respectively.

\subsubsection{Region $V$} First, we show that in Region $V$ the average error
probability does not converge to zero as block length goes to
infinity. This is proved by using a modified version of the
reliability function for rate higher than the channel
capacity~\cite{Dueck_RC}. \vspace{0.1in}
\begin{lemma}{Region $V$}, the average error probability for $X$ does not
converge to $0$ with block length $n$ if $R_x> I(X;Z|Y)$, where
$(X,Y,Z)\sim P_X\times P_Y\times W_{Z|XY}$.
\end{lemma}
\vspace{0.1in} \proof It is enough to show the case where there is
only one message for $Y$ and encoder $Y$ sends a code word $y^n$
with composition $P_Y$. The code book for encoder $X$ is still
uniformly generated among all the fixed-composition-$P_X$ code
books. In the rest of the proof, we investigate the typical behavior
of the codewords $x^n$
 and  modify the Lemma 3 and Lemma 5 from~\cite{Dueck_RC} to show
 that
 \begin{eqnarray}\Pr(\widehat m_x\neq m_x)=P_{e(x)}^n(R_x, R_y, P_X,
 P_Y)>\frac{1}{2}\label{eqn.contra1}
 \end{eqnarray} for large $n$.
 The details of the proof are in Appendix~\ref{section.appendix2}. \hfill $\square$

 \vspace{0.1in}

\subsubsection{Region $IV$}
 The more complicated case is in Region $IV$. We show that the
 decoding error probability for user $X$ does not converge to zero
 with block length $n$. The proof is by contradiction. The idea is to construct
  a decoder that decodes both  message $m_x$ and message $m_y$ correctly with high
  probability, if the decoding error for $m_x$ converges to zero.
 Then again by using a modified proof used in proving
 the reliability function for rate higher than
channel capacity in~\cite{Dueck_RC}, we get a contradiction.

\vspace{0.1in}
\begin{lemma}{Region $IV$}, the average error probability for $X$ does not
converge to $0$ with block length $n$ if $R_x< I(X;Z|Y)$, $R_y<
I(Y;Z|X)$ and $R_x+R_y> I(X,Y;Z)$ where $(X,Y,Z)\sim P_X\times
P_Y\times W_{Z|XY}$.\label{lemma.regionIV}
\end{lemma}

\vspace{0.1in}
 \proof Suppose that
\begin{eqnarray}
 \Pr(\widehat m_x\neq
m_x)=P_{e(x)}^n(R_x, R_y, P_X, P_Y)\leq
\delta_n\label{eqn.decoder3.0}
\end{eqnarray} where $\delta_n$ goes to zero with $n$. Let decoder
$X$ decode $m_y$ by the same decoding rule devised
in~(\ref{eqn.decoder1}):
\begin{eqnarray}
\widehat m_y(z^n)=\argmax_{j\in\{1,2,..., 2^{nR_y}\}} I(z^n;
x^n(\widehat m_x(z^n)), y^n(j)).\label{eqn.decoder3}
\end{eqnarray}
The decoding error for either message at decoder $X$ is now:
\begin{eqnarray}
\Pr((\widehat m_x, \widehat m_y)\neq (m_x,m_y))&=& \Pr( \widehat m_x \neq  m_x )+\Pr( \widehat m_x=m_x , \widehat m_y \neq  m_y)  \nonumber\\
&\leq &\Pr( \widehat m_x \neq  m_x )+\Pr( \widehat m_y \neq
m_y|\widehat m_x=m_x)\label{eqn.conditional_error}
\end{eqnarray}
Given $\widehat m_x=m_x$, (\ref{eqn.decoder3}) becomes
\begin{eqnarray}
\widehat m_y(z^n)=\argmax_{j\in\{1,2,..., 2^{nR_y}\} } I(z^n;
x^n(m_x ), y^n(j)).\label{eqn.decoder3a}
\end{eqnarray}
So the second term in the RHS of~(\ref{eqn.conditional_error}),
$\Pr( \widehat m_y \neq m_y|\widehat m_x=m_x)$,  can be upper
bounded as shown in~(\ref{eqn.proofpart2}). Substitute the upper
bounds~(\ref{eqn.proofpart2}) and~(\ref{eqn.decoder3.0}) into
(\ref{eqn.conditional_error}),  we have:
\begin{eqnarray}
\Pr((\widehat m_x, \widehat m_y)\neq (m_x,m_y)) \leq
\delta_n+2^{-n(E_{y|x}-\epsilon_n)}\label{eqn.contra2a}
\end{eqnarray}
This upper bound~(\ref{eqn.contra2a}) converges to $0$ as $n$ goes
to infinity. However in Appendix~\ref{section.appendix2}, we show
that
 \begin{eqnarray}P_{e(xy)}^n(R_x, R_y, P_X,
 P_Y)=\Pr((\widehat m_x, \widehat m_y)\neq (m_x,m_y))>\frac{1}{2}\label{eqn.contra2}
 \end{eqnarray}
This is contradicted to~(\ref{eqn.contra2a}).
 \hfill $\square$

\vspace{0.1 in}

\subsubsection{Region $III$} This is  a corollary of Lemma~\ref{lemma.regionIV}.
This is intuitively  obvious since for each rate pair $(R_x, R_y)$
in Region $III$, we can find a rate pair $(R_x, R'_y)$ in Region
$IV$ such that $R_y> R'_y$. We construct a contradiction as follows.
For a $(R_x,R_y)$ decoder,  we can construct a new decoder for
$(R_x,R'_y)$ where $R'_y< R_y$, by revealing a random selection of a
$(R_x,R_y)$ code book that is the superset of the $(R_x,R'_y)$ code
book to the $(R_x,R_y)$ decoder and accept the estimate of the
$(R_x,R_y)$ decoder as the estimate for the $(R_x,R'_y)$ decoder. If
the average error probability is small for the $(R_x,R_y)$ code
books, the average error probability is small for this particular
$(R_x,R'_y)$ decoder as well, this is a contradiction to
Lemma~\ref{lemma.regionIV}. Hence the decoding error for encoder $X$
does not converge to $0$ with $n$ if the rate pair $(R_x, R_y)$ is
in Region $III$. \hfill $\square$\vspace{0.1in}

 This concludes the converse part of the
proof for Theorem~\ref{Thm.Int_capacity}.

 \section{Discussions on Time-sharing }\label{sec.timeshare}

The main result of this paper is the randomized fixed-composition
coding capacity region for $X$ that is $\mathcal R_x(P_X, P_Y)$
shown in Figure~\ref{fig.inter_region}. So obviously, the
interference channel capacity region, where  decoding errors for
both $X$ and $Y$ are small, is the intersection of $\mathcal
R_x(P_X, P_Y)$ and $\mathcal R_y(P_X, P_Y)$ where $\mathcal R_y(P_X,
P_Y)$ is defined in the similar way but with channel $\tilde
W_{\tilde Z|XY}$ instead of $W_{Z|XY}$.  The intersected region
defined in~(\ref{eqn.XYREGION}), $\mathcal R_{xy}(P_X, P_Y)$,  is in
general non-convex  as shown in Figure~\ref{fig.inter_region_XY}.
Similar to multiple-access channels capacity region, studied in
Chapter~15.3~\cite{Cover}, we use this capacity region $\mathcal
R_{xy}(P_X, P_Y)$ as the building blocks to generate larger capacity
regions.

\subsection{A digression to MAC channel capacity region}
Before giving the time-sharing results for interference channels and
show why the simple time-sharing idea works for MAC channels but not
for interference channels, we first look at $\mathcal R_x(P_X, P_Y)$
in Figure~\ref{fig.inter_region}. Region $II$ is obviously the
multiple access channel $W_{Z|XY}$ region achieved by input
composition $(P_X, P_Y)$ at the two encoders, denoted by $\mathcal
R_{xy}^{mac}(P_X \times P_Y)$. In~\cite{Cover}, the full description
of the MAC channel capacity region is given in two different
manners:

\begin{eqnarray}
  CONVEX\left(\bigcup_{P_X, P_Y }\mathcal R_{xy}^{mac}(P_X \times P_Y)\right)\nonumber
= CLOSURE \left(\bigcup_{P_U, P_{X|U}, P_{Y|U}}\mathcal
R_{xy}^{mac}(P_{X|U} \times P_{Y|U}\times
P_U)\right)\label{eqn.mac_equi}
\end{eqnarray}
where $R_{xy}^{mac}(P_{X|U} \times P_{Y|U}\times P_U)=\{(R_x, R_y):
R_x\leq I(X;Z|Y,U), R_y\leq I(Y;Z|X,U),
     R_x+R_y\leq I(X,Y;Z|U)\} $ and $U$ is the time-sharing auxiliary random variable and $|U|=4$.

The LHS of~(\ref{eqn.mac_equi}) is the convex hull of all the
fixed-composition MAC channel capacity regions. The RHS
of~(\ref{eqn.mac_equi}) is the closure (without convexification) of
all the time-sharing MAC capacity regions.%
 The
equivalence in~(\ref{eqn.mac_equi}) is non-trivial, it is not a
consequence of the tightness of the achievable region. It hinges on
the convexity of the ``basic'' capacity regions $\mathcal
R_{xy}^{mac}(P_X, P_Y)$. As will be shown in
Section~\ref{sec.beyond}, this is not the case for interference
channels, i.e.~(\ref{eqn.mac_equi}) does not hold anymore.

\subsection{Simple time-sharing capacity region and error exponent}
The simple idea of time-sharing is well studied  for multi-user
channel coding, broadcast channel coding. Whenever there are two
operational points $(R^1_x, R^1_y), (R^2_x, R^2_y)$, while there
exist two coding schemes to achieve small error probability at each
operational point, one can use $\lambda n$ amount of channel uses at
$(R^1_x, R^1_y)$ with coding scheme $1$ and $(1-\lambda) n$ amount
of channel uses at $(R^2_x, R^2_y)$ with coding scheme $2$. The rate
of this coding scheme is $(\alpha R^1_x+(1-\alpha) R^2_x, \alpha
R^1_y+(1-\alpha) R^2_y)$ and the error probability is still
small\footnote{The error exponent is, however, at most half of the
individual error exponent. } (no bigger than the sum of two small
error probabilities). This idea is easily generalized to more than
$2$ operational points.

This simple time sharing idea works perfectly for MAC channel coding
as shown in~(\ref{eqn.mac_equi}). The whole capacity region can be
described as time sharing among fixed-composition codes where the
fixed-composition codes are building blocks. If we extend this idea
to interference channel, we have the following simple time sharing
region as discussed in Section~\ref{sec.time-sharing-firstrun}:

\begin{eqnarray}
CONVEX\left(\bigcup_{P_X, P_Y} \mathcal R_{xy}(P_X,P_Y)\right)= 
CONVEX\left(\bigcup_{P_X, P_Y} \mathcal R_{x}(P_X,P_Y) \bigcap
R_{y}(P_X,P_Y)\right).\label{eqn.interference_region}
\end{eqnarray}

We shall soon see in the next section that this result can be
improved.

\subsection{Beyond simple time-sharing: ``Uniform'' time-sharing}\label{sec.beyond}

In this section we give a time-sharing coding scheme that was first
developed by Gallager~\cite{Gallager_MAC} and later further studied
 for universal decoding by Pokorny and
Wallmeier~\cite{Pokorny_MAC} to get better error exponents for MAC
channels. This type of ``uniform'' time-sharing schemes not only
achieves better error exponents, more importantly, we show that this
achieve \textbf{ bigger} capacity region than the simple
time-sharing scheme does for interference channels! Unlike the
multiple-access channels where the simple time-sharing achieves the
whole capacity region, this is unique to the interference channels,
due to the fact that the capacity region is the convex hull of the
intersections of pairs of non-convex regions (convex or not is not
the issue here, the real difference is the intersection operation).

The organization of this section parallel to that for the
fixed-composition. We first introduce the ``uniform'' time-sharing
coding scheme, then give the achievable error exponents and lastly
drive the achievable rate region for such coding schemes. The proofs
are omitted since they are similar to those for the
 randomized fixed-composition codes.
\vspace{0.1in}

\begin{definition}{``Uniform''  time-sharing codes}\label{def:randomized_coding_TS}:
 for a probability distribution  $P_U$ on $\cal U$,  where $ \mathcal U =\{u_1, u_2,..., u_K\}$ with $\sum_{i=1}^K P_U(u_i)=1$,
  and a pair of conditional independent distributions $P_{X|U},
  P_{Y|U}$. We define the two codeword sets\footnote{Again, we ignore the nuisance
  of the non-integers here.} as
  $$X_c(n)=\{x^n: x_1^{n P_U(u1)}\in P_{X|u_1},x_{n P_U(u_1)+1}^{n (P_U(u_1)+P_U(u_2))}\in P_{X|u_2} ,..., x_{n (1-P_U(u_1))}^{n }\in P_{X|u_L}
  \}$$
i.e. the $i$'th chunk of the codeword $x^n $ with length $nP_U(u_i)$
has composition $P_{X|u_i}$, and similarly
$$Y_c(n)=\{y^n: y_1^{n
P_U(u1)}\in P_{Y|u_1},y_{n P_U(u_1)+1}^{n (P_U(u_1)+P_U(u_2))}\in
P_{Y|u_2} ,..., y_{n (1-P_U(u_1))}^{n }\in P_{Y|u_L} \}.$$  A
``uniform'' time-sharing code $(R_x, R_y, P_U P_{X|U}  P_{Y|U})$
encoder picks a code book with the following probability: for any
message $m_x\in \{1,2,...,2^{nR_x}\}$, the code word $x^n(m_x)$ is
uniformly distributed in $X_c(n)$, similarly for encoder Y.
\end{definition}
\vspace{0.1in}

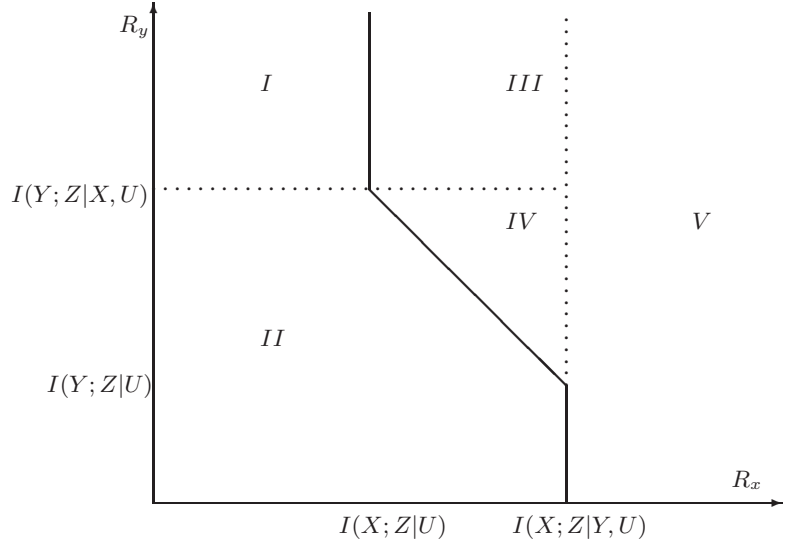
\begin{figure*}
\setlength{\unitlength}{3247sp}%
\begingroup\makeatletter\ifx\SetFigFont\undefined%
\gdef\SetFigFont#1#2#3#4#5{%
  \reset@font\fontsize{#1}{#2pt}%
  \fontfamily{#3}\fontseries{#4}\fontshape{#5}%
  \selectfont}%
\fi\endgroup%
\begin{picture}(5562,4416)(-1701,-4240)
\thinlines { \put(1951,-3961){\vector( 1, 0){4800}}
}%
{ \put(1951,-3961){\vector( 0, 1){3825}}
}%
\thicklines { \put(3601,-211){\line( 0,-1){1350}}
\put(3601,-1561){\line( 1,-1){1500}} \put(5101,-3061){\line(
0,-1){900}}
}%
{
\multiput(5101,-3061)(0.00000,89.9054){32}{\makebox(6.6667,10.0000){\SetFigFont{10}{12}{\rmdefault}{\mddefault}{\updefault}.}}
}%
{
\multiput(1951,-1561)(90.00000,0.00000){35}{\makebox(6.6667,10.0000){\SetFigFont{10}{12}{\rmdefault}{\mddefault}{\updefault}.}}
}%
\put(1626,-361){\makebox(0,0)[lb]{\smash{{\SetFigFont{9}{14.4}{\rmdefault}{\mddefault}{\updefault}{ $R_y$}%
}}}}
\put(6301,-3836){\makebox(0,0)[lb]{\smash{{\SetFigFont{9}{14.4}{\rmdefault}{\mddefault}{\updefault}{ $R_x$}%
}}}}
\put(2701,-811){\makebox(0,0)[lb]{\smash{{\SetFigFont{9}{14.4}{\rmdefault}{\mddefault}{\updefault}{ $I$}%
}}}}
\put(2701,-2761){\makebox(0,0)[lb]{\smash{{\SetFigFont{9}{14.4}{\rmdefault}{\mddefault}{\updefault}{ $II$}%
}}}}
\put(4576,-811){\makebox(0,0)[lb]{\smash{{\SetFigFont{9}{14.4}{\rmdefault}{\mddefault}{\updefault}{ $III$}%
}}}}
\put(4576,-1861){\makebox(0,0)[lb]{\smash{{\SetFigFont{9}{14.4}{\rmdefault}{\mddefault}{\updefault}{ $IV$}%
}}}}
\put(6001,-1861){\makebox(0,0)[lb]{\smash{{\SetFigFont{9}{14.4}{\rmdefault}{\mddefault}{\updefault}{ $V$}%
}}}}
\put(3301,-4186){\makebox(0,0)[lb]{\smash{{\SetFigFont{9}{14.4}{\rmdefault}{\mddefault}{\updefault}{ $I(X;Z|U)$}%
}}}}
\put(1076,-3111){\makebox(0,0)[lb]{\smash{{\SetFigFont{9}{14.4}{\rmdefault}{\mddefault}{\updefault}{ $I(Y;Z|U)$}%
}}}}
\put(801,-1661){\makebox(0,0)[lb]{\smash{{\SetFigFont{9}{14.4}{\rmdefault}{\mddefault}{\updefault}{ $I(Y;Z|X,U)$}%
}}}}
\put(4626,-4186){\makebox(0,0)[lb]{\smash{{\SetFigFont{9}{14.4}{\rmdefault}{\mddefault}{\updefault}{ $I(X;Z|Y,U)$}%
}}}}
\end{picture}%

\caption[ ]{``Uniform'' time-sharing capacity region $\mathcal
R_x(P_U P_{X|U} P_{Y_U})$ for $X$, the achievable region is the
union of Region $I$ and $II$. This region is very similar to that
for fixed-composition coding shown in Figure~\ref{fig.inter_region},
only difference is now there is an auxiliary time-sharing random
variable $U$.}
    \label{fig.inter_region_TIMESHARE}
\end{figure*}

After the code book is randomly generated and revealed to the
decoder, the decoder uses a maximum mutual information decoding
rule. Similar to the fixed-composition coding, the decoder needs to
either decode both message $X$ and $Y$ jointly or simply treats $Y$
as noise and decode $X$ only, depending on where the rate pairs are
in Region $I$ or $II$, as shown in
Figure~\ref{fig.inter_region_TIMESHARE}. The error probability we
investigate is again the average error probability over all messages
and code books.

\vspace{0.1in}

\begin{theorem}{Interference channel capacity region $\mathcal R_{x}(P_U P_{X|U}P_{Y|U})$ for
 ``uniform'' time-sharing codes with composition $P_U P_{X|U}P_{Y|U}$:}
\label{Thm.Int_capacity_time_share}
\begin{eqnarray}\label{eqn.int_region_timesharing}
\mathcal R_x(P_U
P_{X|U}P_{Y|U}) &=&\{(R_x,R_y): 0\leq R_x< I(X;Z|U), 0\leq R_y\}\ \ \ \bigcup\nonumber\\
&& \{(R_x,R_y): 0\leq R_x< I(X;Z|Y,U),  R_x+R_y< I(X,Y;Z|U)\}
\end{eqnarray}
where the random variables in (\ref{eqn.int_region_timesharing}),
$(U, X, Y, Z)\sim P_U P_{X|U} P_{Y|U} W_{Z|X,Y}$. And the
interference capacity region for $P_U P_{X|U}P_{Y|U}$ is
\begin{eqnarray}
\mathcal R_{xy}(P_U P_{X|U}P_{Y|U})=\mathcal  R_{x}(P_U
P_{X|U}P_{Y|U})\bigcap  \mathcal  R_{y}(P_U
P_{X|U}P_{Y|U})\label{eqn.simple-sharing-region-fina}
\end{eqnarray}

\end{theorem}
\vspace{0.1in}

 The rate region defined in
(\ref{eqn.int_region_timesharing}) itself does not give any new
$X$-capacity regions for $X$, since both Region $I$ and $II$ in
Figure~\ref{fig.inter_region_TIMESHARE} can be achieved by simple
time-sharing of Region $I$ and $II$ repectively
in~(\ref{eqn.int_region}). But for the interference channel
capacity, we argue in the next section that this coding scheme gives
a strictly bigger capacity region than that given by the simple
time-sharing of fixed-composition codes
in~(\ref{eqn.interference_region}).

The proof of Theorem~\ref{Thm.Int_capacity_time_share} is
 similar to that of Theorem~\ref{Thm.Int_capacity}. We omit the
 details here. We only point out that the achievability part is
 proved by deriving a positive error exponent for rate pair in the
 interior of the capacity region defined in
 Theorem~\ref{Thm.Int_capacity_time_share}. As shown
 in~\cite{Pokorny_MAC} and also detailed in this paper for the randomized coding,
 the error exponents in Region $II$ of in
 Figure~\ref{fig.inter_region_TIMESHARE} is:
 $$E=\min \{E_{xy}, E_{x|y}, E_{y|x}\},\mbox{ where}$$
\begin{eqnarray} E_{xy}&=&\min_{Q_{XYZ|U}:Q_{X|U}=P_{X|U}, Q_{Y|U}=P_{Y|U}}\nonumber\\
&& D(Q_{Z|XY}\|W|Q_{XYU})
+D(Q_{XY|U}\|P_{X|U}\times P_{Y|U}|U)   +|I_Q(X,Y;Z)-R_x-R_y|^+\nonumber\\
E_{x|y}&=&\min_{Q_{XYZ|U}:Q_{X|U}=P_{X|U},
Q_{Y|U}=P_{Y|U}}\nonumber\\&& D(Q_{Z|XY}\|W| Q_{XYU})
 + D(Q_{XY|U}\|P_{X|U}\times P_{Y|U}|U)  +|I_Q(X;Z|Y,U)-R_x|^+\nonumber\\
E_{y|x}&=&\min_{Q_{XYZ|U}:Q_{X|U}=P_{X|U},
Q_{Y|U}=P_{Y|U}}\nonumber\\&& D(Q_{Z|XY}\|W| Q_{XYU}) +
D(Q_{XY|U}\|P_{X|U}\times P_{Y|U}|U)
   +|I_Q(Y;Z|X,U)-R_y|^+\nonumber
\end{eqnarray}

This is the error exponents  in Lemma~\ref{lemma:regionII} with a
conditional auxiliary random variable $U$.

The error exponent in Region $I$ is
\begin{eqnarray}
E_{x}= && \min_{Q_{XYZ|U}:Q_{X|U}=P_{X|U}, Q_{Y|U}=P_{Y|U}}\nonumber\\
&& D(Q_{Z|XY}\|W| Q_{XYU}) + D(Q_{XY|U}\|P_{X|U}\times P_{Y|U}|U)
+|I_Q(X;Z|U)-R_x|^+\nonumber
\end{eqnarray}

\subsection{Why the ``uniform'' time sharing is needed?}

It is obvious that the ``uniform'' time-sharing fixed-composition
coding gives a bigger error exponent than the simple time-sharing
coding does. More interestingly, we argue that it  gives a bigger
interference channel capacity region. First we write down the
interference channel capacity region generated from the basic
``uniform'' time-sharing fixed-composition codes:
\begin{eqnarray}
CONVEX && \left(\bigcup_{P_{X|U} P_{Y|U} P_U} \mathcal R_{xy}(P_U
P_{X|U}P_{Y|U})\right).\label{eqn.interference_region_TIMESHARING}
\end{eqnarray}
where $ \mathcal R_{xy}(P_U P_{X|U}P_{Y|U})$ is defined
in~(\ref{eqn.simple-sharing-region-fina}) and $CONVEX(A)$ is the
convex hull (simple time sharing) of set $A$.

$U$ is a time-sharing auxiliary random variable. Unlike the MAC
coding problem, where simple time-sharing of fixed-composition codes
achieve the full capacity region, it is not guaranteed for
interference channels. The reason is the intersection operator in
the basic building blocks in~(\ref{eqn.XYREGION})
and~(\ref{eqn.simple-sharing-region-fina}) respectively, i.e. the
interference nature of the problem\footnote{ To understand why
intersection is the difference but not the non-convexity, we
consider four convex sets: $A_1, A_2, B_1, B_2$. We show that
$CONVEX(A_1\bigcap B_1,A_2\bigcap B_2)$ can be strictly smaller than
$CONVEX(A_1, A_2)\bigcap CONVEX (B_1,B_2)$. Let $ A_1=B_2\subset
B_1= A_2$, then $CONVEX(A_1\bigcap B_1,A_2\bigcap B_2)=A_1$ is
strictly smaller than  $CONVEX(A_1, A_2)\bigcap CONVEX
(B_1,B_2)=A_2$. This shows why uniform time-sharing gives bigger
capacity
 region. }.

Obviously the rate region by simple time sharing of fixed
composition code in~(\ref{eqn.interference_region}) is a subset of
simple time sharing of the ``uniform'' time sharing capacity
region~(\ref{eqn.interference_region_TIMESHARING}).  In the
following example, we illustrate
why~(\ref{eqn.interference_region_TIMESHARING})   is
bigger than~(\ref{eqn.interference_region}).\\

\begin{figure}
\begin{center}
\includegraphics[width=90mm]{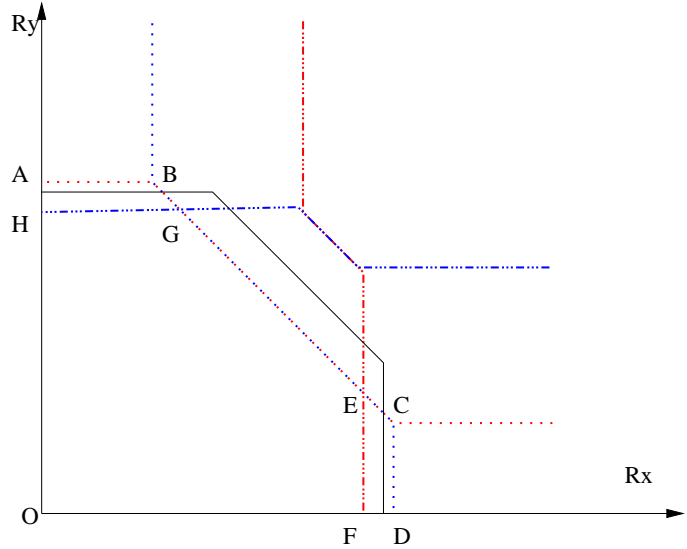}
\end{center}
\caption[ ]{Simple timesharing of fixed-composition capacity $ABCDO$
VS  time-sharing fixed composition  capacity(0.5) ( the black
pentagon)}
    \label{fig.TIMESHARE}
\end{figure}

 \textbf{\textit{Example: }} Suppose we have a symmetric interference channel, i.e. $\mathcal
R_{x}(P_{X} ,P_{Y})= \mathcal R^T_{y}(P_{Y}, P_{X})$  for all $P_X ,
P_Y$ where $^T$ is the transpose operation. The comparison of simple
timesharing capacity region and the more sophisticated time-sharing
fixed-composition capacity region are illustrated by a toy example
in Figure~\ref{fig.TIMESHARE}.

 For a distribution $(P_X, P_Y)$, the
achievable region for the fixed-composition code is illustrated in
Figure~\ref{fig.TIMESHARE}, $\mathcal R_x (P_X, P_Y)$ and $\mathcal
R_y (P_X, P_Y)$ respectively, these are bounded by the red dotted
lines and red dash-dotted lines respectively, so the interference
capacity region $\mathcal R_{xy}(P_X, P_Y)$ is bounded by the
pentagon $ABEFO$. By symmetry, $\mathcal R_x (P_Y, P_X)$ and
$\mathcal R_y (P_X, P_Y)$ are bounded by the blue dotted lines and
blue dash-dotted lines respectively, the capacity region $\mathcal
R_xy(P_Y, P_X)$ is bounded by the pentagon $HGCDO$. So the convex
hull of these two regions is $ABCDO$.

Now consider the following  timesharing fixed-composition coding
$P_{X|U}P_{Y|U}P_U$ where $\mathcal U=\{0,1\}$, $P_U(0)=P_U(1)=0.5$
and $P_{X|0}=P_{Y|1}=P_X$, $P_{X|1}=P_{Y|0}=P_Y$. The interference
capacity region is obviously bounded by the black pentagon
in~Figure~\ref{fig.TIMESHARE}. This toy example shows
why~(\ref{eqn.interference_region_TIMESHARING}) is bigger
than~(\ref{eqn.interference_region}).

\section{Future directions}

The most interesting question about interference channel is the
geometry of the two code books.   For point to point channel coding,
the code words in the optimal code book is uniformly distributed on
a sphere of the optimal compositions and the optimal composition
achieves the capacity. For MAC channels, a simple time-sharing among
different fixed-composition codes is sufficient and necessary to
achieve the whole capacity region, meanwhile for each
fixed-composition codes, the codewords are uniformly distributed.
However as illustrated in Section~\ref{sec.timeshare}, a more
interesting ``uniform'' time sharing is needed. So what is time
sharing? Both simple time sharing and ``uniform'' time sharing
change the shape of the code books, however, in different ways.
Simple time sharing ``glue'' segments of  code words together due to
the independence of the coding in different segments of the channel
uses, meanwhile for ``uniform'' time sharing, code words  still have
equal distances between one another. Better understanding of the
shape of code books may help us understand the interference
channels. Also in this paper, we give our first attempt at giving an
outer bound of the interference channel capacity region. We only
manage to give a tight outer bound to the time-sharing
fixed-composition code. An important future direction is to
categorize the coding schemes for interference channels and more
outer bound result may follow. This is in contrast to the
traditional outer bound derivations~\cite{Carleial} where genie is
used.
%
%
%
%
%
%
%

\section*{Acknowledgments}
The author thanks Raul Etkin, Neri Merhav and Erik Ordentlich  for
introducing the problem and helpful discussions along the way.

\bibliographystyle{plain}

\bibliography{./Interference_Main}

\appendix{}
\subsection{Proof of (\ref{eqn.proofpart1}), (\ref{eqn.proofpart2}) and
(\ref{eqn.inter_error_avg})}\label{section.appendix1}

We give a unified proof in lower bounding the error probability for
randomized fixed-composition coding, where the error probabilities
in (\ref{eqn.proofpart1}), (\ref{eqn.proofpart2}) and
(\ref{eqn.inter_error_avg}) are taken over all messages,  code books
and  channel behaviors. We examine the object function to be
minimized  in (\ref{eqn.proofpart1}), (\ref{eqn.proofpart2}) and
(\ref{eqn.inter_error_avg}).

 First, the \textit{common} part of the
three error exponents $E_{xy}$, $E_{x|y}$ and $E_x$: $
D(Q_{Z|XY}\|W| Q_{XY})+D(Q_{XY}\|P_X\times P_Y)$.
$D(Q_{XY}\|P_X\times P_Y)$ is the logarithm of the inverse of the
probability that   type $Q_{XY}$ is the empirical distribution of
the code pair $x^n(1), y^n(1)$ individually generated from
fixed-compositions $P_X$ and $P_Y$. $ D(Q_{Z|XY}\|W| Q_{XY})$ is
logarithm of the inverse of the conditional probability that the
input to the channel $W$ is $Q_{XY}$, while the empirical type of
the input/output is $Q_{XYZ}= Q_{XY}\times Q_{Z|XY}$.

Secondly for the individual part of the error exponents in
(\ref{eqn.proofpart1}), (\ref{eqn.proofpart2}) and
(\ref{eqn.inter_error_avg}): $|I_Q(X,Y;Z)-R_x-R_y|^+ $,
$|I_Q(X;Z|Y)-R_x|^+$ and $|I_Q(X;Z)-R_x|^+$ respectively, each one
is the logarithm of the inverse of an upper bound on the probability
that there exists another message (pair) with higher mutual
information with the channel output, while  the channel inputs/ouput
has type $Q_{XYZ}$. This is derived by a union bound argument. We
now give the details of the proofs. \vspace{0.1in}

\subsubsection{Proof of~(\ref{eqn.proofpart1})}

Because of the symmetry of the code book selection, we can fix the
message pair $(m_x,m_y) = (1,1)$ and write the error
probability~(\ref{eqn.proofpart1}) in the following way:
\vspace{0.1in}

$\Pr( m_x\neq \widehat m_x, m_y\neq \widehat m_y )$
\begin{eqnarray}
&=&\left (\frac{1}{|\mathcal T^n(P_X)|}\right)^{2^{nR_x}}\left
(\frac{1}{|\mathcal
T^n(P_Y)|}\right)^{2^{nR_y}}\sum_{c_X}\sum_{c_Y}\label{eqn.interp1}\\
&&\hspace{0.1in}\frac{1}{2^{nR_x}}\sum_{m_x}\frac{1}{2^{nR_y}}\sum_{m_y}\sum_{z^n}
  W_{Z|XY}(z^n|x^n(m_x),y^n(m_y)) 1(\widehat m_x(z^n)\neq
m_x, \widehat m_y(z^n)\neq
m_y)\nonumber\\
&=&\left (\frac{1}{|\mathcal T^n(P_X)|}\right)^{2^{nR_x}}\left
(\frac{1}{|\mathcal
T^n(P_Y)|}\right)^{2^{nR_y}}\nonumber\\
 &&\hspace{1.1in} \sum_{c_X}\sum_{c_Y} \sum_{z^n}
  W_{Z|XY}(z^n|x^n(1),y^n(1)) 1(\widehat m_x(z^n)\neq
1,\widehat m_y(z^n)\neq
1 )\nonumber\\
&=&\sum_{Q_{XY}:Q_X=P_X, Q_Y=P_Y} \big\{\Pr \big((x^n(1), y^n(1))\in
Q_{XY} \big)\sum_{Q_{Z|XY}}\Pr(z^n|(x^n(1),
y^n(1))\in Q_{Z|XY})\nonumber\\
  &&\hspace{1.1in}  \Pr(\widehat m_x(z^n)\neq
1,\widehat m_y(z^n)\neq 1 )\big\}\label{eqn.interp2}\\
&\leq&\sum_{Q_{XY}:Q_X=P_X, Q_Y=P_Y} \big\{\Pr \big((x^n(1),
y^n(1))\in Q_{XY} \big)\sum_{Q_{Z|XY}}\Pr(z^n|(x^n(1),
y^n(1))\in Q_{Z|XY})\nonumber\\
 &&  \min\{1, \sum_{i=2}^{2^{nR_x}}\sum_{j=2}^{2^{nR_y}}\Pr\left(I(z^n;x^n(1), y^n(1))\leq I(z^n;x^n(i), y^n(j))
 |(x^n(1), y^n(1), z^n)\in Q_{XYZ})\right)\}\big\}\nonumber\\
&\leq&|\mathcal T^n_{XYZ}|\max_{Q_{XYZ}:Q_X=P_X, Q_Y=P_Y}  \Pr
\big((x^n(1), y^n(1))\in Q_{XY} \big) \Pr(z^n|(x^n(1),
y^n(1))\in Q_{Z|XY})\label{eqn:4parts}\\
 &&  \min\{1, \sum_{i=2}^{2^{nR_x}}\sum_{j=2}^{2^{nR_y}}\Pr\left(I(z^n;x^n(1), y^n(1))\leq I(z^n;x^n(i), y^n(j))
 |(x^n(1), y^n(1), z^n)\in Q_{XYZ})\right)\} \nonumber
\end{eqnarray}
(\ref{eqn.interp1}) and (\ref{eqn.interp2}) are two different
interpretations of the same error probability. In
(\ref{eqn.interp1}), we first randomly pick  a fixed-composition
code book pair $c_X$ and $c_Y$, then sum over the all probabilities
that the output of the channel causes a decoding error for the
chosen code book pair. (\ref{eqn.interp2}) is an equivalent
interpretation of the above error probability because the codewords
for each message is independently generated. We interpret
(\ref{eqn.interp2}) as follows, we first randomly pick a codeword
pair for message $1$ in $X$ and message $1$ in $Y$, then the
codeword pair is transmitted to through the channel. Then we
randomly generate the rest of the code book and investigate the
probability that other message pairs maximize the mutual information
with the channel output. We upper bound the four terms in
(\ref{eqn:4parts}) individually in (\ref{eqn.4parts1}),
(\ref{eqn.4parts2}), (\ref{eqn.4parts3}) and (\ref{eqn.4parts4}).

First, the number of type sets of length $n$:
\begin{eqnarray}
|\mathcal T^n_{XYZ}|\leq(n+1)^{|\mathcal X \times \mathcal Y \times
 \mathcal Z|} =  2^{n (\frac{\log(n+1)}{n}|\mathcal X \times \mathcal Y \times
 \mathcal Z|)}=2^{n a_n}\label{eqn.4parts1}.
\end{eqnarray}
Secondly, for any $Q_{XY}$, s.t. $Q_X=P_X$ and $Q_Y=P_Y$, from the
method of types~\cite{Cover} and~\cite{Csiszar:98}, we know that $
2^{n (H(P_Y)-\frac{\log n}{n}|\mathcal Y|)}\leq |\mathcal P_Y|\leq
2^{n H(P_Y)} $, similar bounds applies to $|\mathcal P_X|$. And for
a fixed $X$-sequence, $x^n(1)\in P_X=Q_X$, we have
$2^{n(H(Q_{Y|X})-\frac{\log n}{n}|\mathcal XY|)}\leq |\{y^n\in
\mathcal Y^n: (x^n(1), y^n)\in Q_{XY}\}|\leq 2^{nH(Q_{Y|X})}$.
$x^n(1)$ and $y^n(1)$ are independently distributed in type set
$P_X$ and $P_Y$. Hence,
\begin{eqnarray}
\Pr \big((x^n(1), y^n(1))\in Q_{XY} \big)=\frac{ |\{y^n\in \mathcal
Y^n: (x^n(1), y^n)\in Q_{XY}\}|}{|P_Y|}\leq 2^{n
(H(Q_{Y|X})-H(Q_Y)+\frac{\log n}{n}|\mathcal X|)}\nonumber
\end{eqnarray}

Notice that $H(Q_{Y|X})-H(Q_Y)=-D(Q_{XY}\|Q_X\times Q_Y)=-D
(Q_{XY}\|P_X\times P_Y)$ and let $b_n=\frac{\log n}{n}|\mathcal X|$,
we have:
\begin{eqnarray}
\Pr \big((x^n(1), y^n(1))\in Q_{XY} \big)  \leq 2^{-n(D
(Q_{XY}\|P_X\times P_Y)-b_n)} \label{eqn.4parts2}
\end{eqnarray}

Thirdly, For $(x^n(1), y^n(1))\in Q_{XY}$, for any empirical channel
behavior $Q_{Z|XY}$:
\begin{eqnarray}
\Pr(z^n|(x^n(1), y^n(1))\in Q_{Z|XY})&= &|\{z^n: (x^n(1),
y^n(1),z^n)\in Q_{XYZ} \}| W_{Z|XY}(Q_{Z|XY})
\nonumber\\
&\leq & 2^{n H(Q_{Z|XY})}\times
2^{n(-D(Q_{Z|XY}\|W|Q_{XY})-H(Q_{Z|XY}))}\nonumber\\
&=& 2^{-n D(Q_{Z|XY}\|W|Q_{XY})}\label{eqn.4parts3}
\end{eqnarray}

Finally, for $(x^n(1), y^n(1), z^n)\in Q_{XYZ}$, we investigate the
probability that there exists $(i,j)$, $i\neq 1, j\neq 1$, s.t. the
mutual information between $(x^n(i), y^n(j))$ and $z^n$ is at least
as much as the mutual information between $(x^n(1), y^n(1))$ and
$z^n$. For all $i\neq 1$, the codeword $x^n(i)$ is uniformly
distributed on the fixed-composition set $P_X$, same for $Y$. Given
$(x^n(1), y^n(1), z^n)\in Q_{XYZ}$, we have $I(z^n;x^n(1),
y^n(1))=I_Q(Z;X,Y)$, so:

\begin{eqnarray}
 && \min\{1, \sum_{i=2}^{2^{nR_x}}\sum_{j=2}^{2^{nR_y}}\Pr\left(I(z^n;x^n(1), y^n(1))\leq I(z^n;x^n(i), y^n(j))
 |(x^n(1), y^n(1), z^n)\in Q_{XYZ}\right)\}\nonumber\\
 &&\leq \min\{1, 2^{n(R_x+R_y)}\sum_{V_{XYZ}: V_X=Q_X, V_Y=Q_Y, V_Z=Q_Z, I_Q(Z;X,Y)\leq I_V(Z;X,Y)}
\nonumber\\
&&\hspace{2.1in}\Pr(((x^n(i), y^n(j), z^n)\in V_{XYZ}|z^n\in Q_Z) \}
\nonumber\\
 &&= \min\{1, 2^{n(R_x+R_y)}\sum_{V_{XYZ}: V_X=Q_X, V_Y=Q_Y,
V_Z=Q_Z, I_Q(Z;X,Y)\leq I_V(Z;X,Y)}\nonumber\\
&&\hspace{2.1in} \frac{|\{(x^n ,y^n)\in P_X\times P_Y: (x^n
,y^n,z^n)\in V_{XYZ}\}|}{|\{x^n:x^n\in P_X\}||\{y^n:y^n\in P_Y\}|}\}
\nonumber\\
 &&\leq \min\{1, 2^{n(R_x+R_y)}\sum_{V_{XYZ}: V_X=Q_X, V_Y=Q_Y,
V_Z=Q_Z, I_Q(Z;X,Y)\leq I_V(Z;X,Y)}\nonumber\\
&&\hspace{2.1in}2^{n (H_V(X,Y|Z)-H_V(X)-H_V(Y)+\frac{\log n
(|\mathcal X|+|\mathcal Y|)}{n})}\}
\nonumber\\
 &&\leq \min\{1, 2^{n(R_x+R_y)}\sum_{V_{XYZ}: V_X=Q_X, V_Y=Q_Y,
V_Z=Q_Z, I_Q(Z;X,Y)\leq I_V(Z;X,Y)}\nonumber\\
&&\hspace{2.1in}2^{n (H_V(X,Y|Z)-H_V(X,Y)+\frac{\log n (|\mathcal
X|+|\mathcal Y|)}{n})}\}
\nonumber\\
 &&= \min\{1, 2^{n(R_x+R_y)}\sum_{V_{XYZ}: V_X=Q_X, V_Y=Q_Y,
V_Z=Q_Z, I_Q(Z;X,Y)\leq I_V(Z;X,Y)}2^{n (-I_V(X,Y;Z)+\frac{\log n
(|\mathcal X|+|\mathcal
Y|)}{n})}\} \nonumber\\
 &&\leq \min\{1, 2^{n(R_x+R_y)} n^{|\mathcal X \times\mathcal Y\times \mathcal Z|}2^{n (-I_Q(X,Y;Z)+\frac{\log n
(|\mathcal X|+|\mathcal
Y|)}{n})}\} \nonumber\\
 &&= 2^{-n(|I_Q(X,Y;Z)-R_x-R_y |^+-c_n)}\label{eqn.4parts4}
\end{eqnarray}
Substituting (\ref{eqn.4parts1}), (\ref{eqn.4parts2}),
(\ref{eqn.4parts3}) and (\ref{eqn.4parts4}) in~(\ref{eqn:4parts}),
and noticing that $a_n$ $b_n$ and $c_n$ converges to zero when $n$
goes to infinity,   (\ref{eqn.proofpart1}) is proved.

\vspace{0.1in}

\subsubsection{Sketch of the proof of~(\ref{eqn.proofpart2}) and~(\ref{eqn.inter_error_avg})}

(\ref{eqn.proofpart2}) and (\ref{eqn.inter_error_avg}) can be proved
by following the same argument in proving~(\ref{eqn.proofpart1}).
Similar to how we upper bound the LHS of (\ref{eqn.proofpart1}) in
(\ref{eqn:4parts}), we upper bound the LHS of (\ref{eqn.proofpart2})
by: \vspace{0.1in}

 $\Pr( m_x\neq \widehat m_x|
m_y=  \widehat m_y )$
\begin{eqnarray}
&\leq&|\mathcal T^n_{XYZ}|\max_{Q_{XYZ}:Q_X=P_X, Q_Y=P_Y}  \Pr
\big((x^n(1), y^n(1))\in Q_{XY} \big) \Pr(z^n|(x^n(1),
y^n(1))\in Q_{Z|XY})\nonumber\\
 &&  \min\{1, \sum_{i=2}^{2^{nR_x}} \Pr\left(I(z^n;x^n(1), y^n(1))\leq I(z^n;x^n(i), y^n(1))
 |(x^n(1), y^n(1), z^n)\in Q_{XYZ})\right)\}.\label{eqn:4parts_case2}
\end{eqnarray}
and the LHS of (\ref{eqn.inter_error_avg}) by

\vspace{0.1in}

 $\Pr( m_x\neq \widehat m_x)$
\begin{eqnarray}
&\leq&|\mathcal T^n_{XYZ}|\max_{Q_{XYZ}:Q_X=P_X, Q_Y=P_Y}  \Pr
\big((x^n(1), y^n(1))\in Q_{XY} \big) \Pr(z^n|(x^n(1),
y^n(1))\in Q_{Z|XY})\nonumber\\
 &&  \min\{1, \sum_{i=2}^{2^{nR_x}} \Pr\left(I(z^n;x^n(1))\leq I(z^n;x^n(i))
 |(x^n(1), y^n(1), z^n)\in Q_{XYZ})\right)\}.\label{eqn:4parts_case3}
\end{eqnarray}

The common parts (the three terms on the first line) in
(\ref{eqn:4parts_case2}) and (\ref{eqn:4parts_case3}) are upper
bounded the same way as those in (\ref{eqn.4parts1})
(\ref{eqn.4parts2}) and (\ref{eqn.4parts3}) for~(\ref{eqn:4parts}).
 The individual part (the $\min\{1, \cdot\}$ term on the second line)
  of (\ref{eqn:4parts_case2}) and
(\ref{eqn:4parts_case3})  are upper bounded by a similar argument
for upper bounding the individual part of (\ref{eqn:4parts}) shown
in
 (\ref{eqn.4parts3}). We omit the details here. \hfill $\square$
%
%
%


\subsection{Proof of  (\ref{eqn.contra1}) and (\ref{eqn.contra2})}\label{section.appendix2}
We give a constant lower bound, $\frac{1}{2}$, on the error
probabilities $\Pr(\widehat m_x\neq m_x ) $ and $\Pr((\widehat m_x,
\widehat m_y) \neq (m_x, m_y) ) $
 in~(\ref{eqn.contra1}) and (\ref{eqn.contra2}) respectively.  The
technical details  of lower bounding $\Pr(\widehat m_x\neq m_x ) $
is carried out in Appendix~\ref{sec.appendix.detailforX}. We extend
the two very technical Lemmas 5 and 3 from~\cite{Dueck_RC} into
Lemmas~\ref{lemma:Dueck5} and~\ref{lemma:Dueck3} respectively, where
Lemma~\ref{lemma:Dueck3} is used to prove Lemma~\ref{lemma:Dueck5}.
 The proof of lower bounding $\Pr((\widehat m_x, \widehat m_y) \neq
(m_x, m_y) ) $ is similar, we only give the necessary  definition of
jointly good code books  in Appendix~\ref{sec.appendix.detailforXY}.

The difference between the setups in this paper and that
in~\cite{Dueck_RC} is that we are dealing with an interference
channel instead of a memoryless channel in~\cite{Dueck_RC}. Hence a
notion of the conditionally typical code book in the proof
of~(\ref{eqn.contra1}) and jointly typical code book in the proof
of~(\ref{eqn.contra2}) is necessary in the proofs.

\vspace{0.1in}
\subsubsection{Proof of~(\ref{eqn.contra1})}\label{sec.appendix.detailforX}
   we  give an upper bound of the \textit{correct
decoding probability} $\Pr( \widehat m_x =  m_x )=1-\Pr( \widehat
m_x \neq  m_x )$ and hence prove the lower bound on $\Pr( \widehat
m_x \neq  m_x )$ in~(\ref{eqn.contra1}) .
\begin{eqnarray}
\Pr(\widehat m_x= m_x ) &=& P_{e(x)}^n(R_x, R_y, P_X,
 P_Y)\nonumber\\
 &=&\left (\frac{1}{|\mathcal T^n(P_X)|}\right)^{2^{nR_x}} \sum_{c_X}
 \frac{1}{2^{nR_x}}\sum_{m_x} \sum_{z^n}
  W_{Z|XY}(z^n|x^n(m_x),y^n) 1(\widehat m_x(z^n)=
m_x)\nonumber
\end{eqnarray}
The codewords $x^n(m_x)$ is uniformly distributed on the type set
$P_X$, so the probability that the joint type of $(x^n(m_x), y^n)$
is close to $P_X\times P_Y$ with high probability~\cite{Cover}, i.e.
for all $\sigma>0$, for large $n$,
\begin{eqnarray}\label{eqn.typical_codebook}
\Pr(D((x^n(m_x), y^n)\|P_X\times P_Y))> \sigma) <\sigma.
\end{eqnarray}
 We denote
by $T_\sigma(y^n)=\{x^n: D((x^n , y^n)\|P_X\times P_Y))\leq
\sigma\}$, the typical set conditional on $y^n$. We say a code book
$c_X$ is good conditional on $y^n$ if
\begin{eqnarray}\label{eqn:good_def}
|c_X\bigcap T^C_\sigma(y^n)|\leq \frac{|c_X|}{4}
\end{eqnarray}
where $|c_X|=2^{nR_x}$. The set of all good code books is denoted by
$G$, at most $ 4 \sigma$ of the code books are not in $G$ because of
(\ref{eqn.typical_codebook}). For a good code book $c_X$, we use the
technique from~\cite{Dueck_RC} to upper bound the correct
probability for the good code book $c_X$.
\begin{eqnarray}
\Pr(\widehat m_x= m_x)&\leq& \frac{|c_X\bigcap
T^C_\sigma(y^n)|}{|c_X|}+\frac{1}{|c_X|}\sum_{i: x^n(i)
\in T_\sigma(y^n)}\Pr(i= \widehat m_x(z^n))\nonumber\\
&\leq & \frac{1}{4}+\frac{1}{|c_X|}\sum_{i: x^n(i)\in
T_\sigma(y^n)}\Pr(i= \widehat m_x(z^n))\nonumber\\
&\leq & \frac{1}{4}+ 2^{-n(E-\epsilon_n)}\label{eqn.appendix_last}
\end{eqnarray}
 where $\epsilon_n$ goes to zero with $n$, and
\begin{eqnarray}
E=\min_{Q_{XYZ}: D(Q_{XY}\|P_X\times P_Y)< \sigma}
D(Q_{Z|XY}\|W_{Z|XY}|Q_{XY}) +|R_x-I_Q(X;Z|Y)|^+\nonumber
\end{eqnarray}
where (\ref{eqn.appendix_last}) is proved by
Lemma~\ref{lemma:Dueck5} which is  an extension  of Lemma  5
in~\cite{Dueck_RC} from memoryless to conditional on $y^n$.

Following the argument in Lemma~\ref{lemma:positiveness}, it is easy
to see that $E>0$ for $R_x>I(X;Z|Y)$ and small $\sigma$, where
$(X,Y,Z)\sim W_{Z|XY}\times P_X\times P_Y$. Now we have
\begin{eqnarray}
\Pr(\widehat m_x= m_x ) &=&  \left (\frac{1}{|\mathcal
T^n(P_X)|}\right)^{2^{nR_x}} (\sum_{c_X\in G}\Pr(\widehat
m_x=m_x)+\sum_{c_X\in G^C}\Pr(\widehat m_x=m_x))\nonumber\\
&\leq & \frac{1}{4}+ 2^{-n(E-\epsilon_n)}+4\sigma
\end{eqnarray}
Let $\sigma $ be small enough and let $n$ goes to infinity, so
$\Pr(\widehat m_x\neq m_x )=1-\Pr(\widehat m_x= m_x ) \geq
\frac{1}{2} $. (\ref{eqn.contra1}) is proved.\hfill $\square$

\vspace{0.1in}

The following two Lemmas~\ref{lemma:Dueck5} and~\ref{lemma:Dueck3}
are extensions of Lemma 5 and 3 in~\cite{Dueck_RC} respectively.
They contain the technical details in the proof
of~(\ref{eqn.appendix_last}).

  \vspace{0.1in}

\begin{lemma}{Extension of Lemma $5$ in~\cite{Dueck_RC} from memoryless to conditional  on
$y^n$}\label{lemma:Dueck5}, for a good code book $c_X\in G$ defined
in~(\ref{eqn:good_def}). Recall that $|c_X\bigcap T_\sigma(y^n)|\geq
\frac{3|c_X|}{4}= \frac{3}{4}\times 2^{nR_x}$, then for any decoding
rule (previously known as $\widehat m_x$) $\phi :\mathcal
Z^n\rightarrow \{1,2,...,2^{nR_x}\}$,
\begin{eqnarray}\label{eqn.upperbound_correct}
\frac{1}{|c_X|}\sum_{i: x^n(i)\in T_\sigma(y^n)}\Pr(i= \phi(z^n))
\leq  2^{-n(E-\epsilon_n)}
\end{eqnarray}
\begin{eqnarray}
\mbox{where } E=\min_{Q_{XYZ}: D(Q_{XY}\|P_X\times P_Y)< \sigma}
D(Q_{Z|XY}\|W_{Z|XY}|Q_{XY}) +|R_x-I_Q(X;Z|Y)|^+\nonumber
\end{eqnarray}
and $\epsilon_n =\epsilon(|\mathcal X|,|\mathcal Y|,|\mathcal Z|,n)$
which converges to zero as $n$ goes to infinity.
\end{lemma}\vspace{0.1in}
\proof We write $M=\{i\in \{1,2,...,2^{nR_x}\}: x^n(i)\in
T_\sigma(y^n)\}$ then we know that from the definition of a good
code book: $\frac{3}{4}\times 2^{nR_x}\leq |M|\leq 2^{nR_x}=|c_X|$.
Notice that
\begin{eqnarray}\Pr(i=  \phi( z^n))=\sum_{z^n\in \phi^{-1} (i)}
W_{Z|XY}( z^n| x^n(i), y^n)= W_{Z|XY}( \phi^{-1}(i)| x^n(i), y^n)
\end{eqnarray}
We rewrite the LHS of~(\ref{eqn.upperbound_correct}):
\begin{eqnarray}
 &= &2^{-nR_x}\sum_{i:x^n(i)\in T_\sigma(y^n)}W_{Z|XY}( \phi^{-1}(i)| x^n(i), y^n)\nonumber\\
&= & 2^{-nR_x}\sum_{Q_{XY}:D(Q_{XY}\|P_X\times P_Y)< \sigma }
\left(\sum_{i: (x^n(i),y^n)\in Q_{XY}}W_{Z|XY}( \phi^{-1}(i)| x^n(i), y^n)\right)\nonumber\\
&\leq & (n+1)^{|\mathcal X||\mathcal Y|}
\max_{Q_{XY}:D(Q_{XY}\|P_X\times
P_Y)< \sigma } \left(2^{-nR_x}\sum_{i: (x^n(i),y^n)\in Q_{XY}}W_{Z|XY}( \phi^{-1}(i)| x^n(i), y^n)\right)\nonumber\\
&= & (n+1)^{|\mathcal X||\mathcal Y|}
\max_{Q_{XY}:D(Q_{XY}\|P_X\times P_Y)< \sigma }
\nonumber\\
&&\left(2^{-nR_x}\sum_{i: (x^n(i),y^n)\in
Q_{XY}}\sum_{Q_{Z|XY}}W_{Z|XY}( \phi^{-1}(i)\bigcap Q_{Z|XY}(x^n(i),
y^n) | x^n(i), y^n)\right)\nonumber\\
&\leq  & (n+1)^{|\mathcal X||\mathcal Y|+|\mathcal X||\mathcal Y|
|\mathcal Z| } \max_{Q_{XYZ}:D(Q_{XY}\|P_X\times P_Y)< \sigma
}\nonumber\\
&& \left(2^{-nR_x}\sum_{i: (x^n(i),y^n)\in Q_{XY}} W_{Z|XY}(
\phi^{-1}(i)\bigcap Q_{Z|XY}(x^n(i), y^n) | x^n(i),
y^n)\right)\nonumber
\end{eqnarray}

\begin{eqnarray}
&\leq  & 2^{n \epsilon_n(1)} \max_{Q_{XYZ}:D(Q_{XY}\|P_X\times P_Y)<
\sigma }  \nonumber\\
&& \left(2^{-nR_x}\sum_{i: (x^n(i),y^n)\in Q_{XY}} W_{Z|XY}(
Q_{Z|XY}(x^n(i), y^n)| x^n(i), y^n) \frac{|Q_{Z|XY}(x^n(i),
y^n)\bigcap \phi^{-1}(i)|}{|Q_{Z|XY}(x^n(i), y^n)|}\right)\nonumber\\
&\leq  & 2^{n \epsilon_n(1)} \max_{Q_{XYZ}:D(Q_{XY}\|P_X\times P_Y)<
\sigma } \nonumber\\
&&\left(2^{-n D(Q_{Z|XY}\|W_{Z|XY}|Q_{XY})}2^{-nR_x}\sum_{i:
(x^n(i),y^n)\in Q_{XY}}
 \frac{|Q_{Z|XY}(x^n(i),
y^n)\bigcap \phi^{-1}(i)|}{|Q_{Z|XY}(x^n(i), y^n)|}\right)\nonumber\\
&\leq  & 2^{n \epsilon_n(1)} \max_{Q_{XYZ}:D(Q_{XY}\|P_X\times P_Y)<
\sigma } \left(2^{-n D(Q_{Z|XY}\|W_{Z|XY}|Q_{XY})}  2^{-n |R  -
I_Q(X;Z|Y)- \epsilon_n(2)|^+}\right)\label{eqn:lemmalemma}\\
&=& 2^{-n (E-\epsilon_n)}
\end{eqnarray}
where~(\ref{eqn:lemmalemma}) follows Lemma~\ref{lemma:Dueck3}. The
rest are obvious by the method of types. \hfill $\square$
\vspace{0.1in}

\vspace{0.1in}
 \begin{lemma}{Extension of Lemma $3$
in~\cite{Dueck_RC} from memoryless to conditional on
$y^n$,}\label{lemma:Dueck3} for any $R\geq R_x>0$, for any coding
system $X(y^n)$ with joint input distribution $(x^n(i), y^n)\in
Q_{XY}$, $i=1,2,... 2^{nR_x}$, and decoding rule $\phi :\mathcal
Z^n\rightarrow \{1,2,...,2^{nR_x}\}$, let  $Q_{Z|XY}(x^n(i),
y^n)=\{z^n: (x^n(i), y^n, z^n) \in Q_{XYZ}\}$ (this is the V-shell
notation $T_V$ used in~\cite{Dueck_RC}),  we have:
\begin{eqnarray}\label{eqn.appendix1}
\frac{1}{2^{nR}} \sum_{i=1}^{2^{nR_x}} \frac{|Q_{Z|XY}(x^n(i),
y^n)\bigcap \phi^{-1}(i)|}{|Q_{Z|XY}(x^n(i), y^n)|}\leq 2^{-n |R  -
I_Q(X;Z|Y)- \epsilon_n|^+}
\end{eqnarray}
where $\epsilon_n=\epsilon(n, |\mathcal X|,|\mathcal  Y|,|\mathcal
 Z|)$ converges to zero as $n$ goes to infinity.

\end{lemma}\vspace{0.1in}
 \proof Write $Q_{Z|Y}( y^n)=\{z^n: (y^n, z^n)\in Q_{ZY}\}$. By the method of types~\cite{Csiszar:98}, we know that
\begin{eqnarray*}
(n+1)^{-|\mathcal Z|}2^{ n H_Q(Z|XY)}\leq  |Q_{Z|XY}(x^n(i), y^n)|
\leq 2^{ n H_Q(Z|XY)}
\end{eqnarray*}
\begin{eqnarray*}
\mbox {and }  (n+1)^{-|\mathcal Z|}2^{ n H_Q(Z|Y)}\leq  |Q_{Z|Y}(
y^n)| \leq 2^{ n H_Q(Z|Y)}.
\end{eqnarray*}

So the LHS of (\ref{eqn.appendix1}) is upper bounded by
\begin{eqnarray}
&&\frac{1}{2^{nR }} \sum_{i=1}^{2^{nR_x}} \frac{|Q_{Z|XY}(x^n(i),
y^n)\bigcap \phi^{-1}(i)|}{|Q_{Z|XY}(x^n(i), y^n)|}\nonumber\\
&\leq& (n+1)^{ |\mathcal Z|}2^{- n H_Q(Z|XY)}2^{-n R}
\sum_{i=1}^{2^{nR_x }}
 |Q_{Z|XY}(x^n(i), y^n)\bigcap \phi^{-1}(i)|\nonumber\\
&\leq & (n+1)^{ |\mathcal Z|}2^{- n H_Q(Z|XY)}2^{-n R }
  |Q_{Z|Y}( y^n)| \label{eqn.appendix2}   \\
&\leq & (n+1)^{ |\mathcal Z|}2^{ -n H_Q(Z|XY)}2^{-n R }
  (n+1)^{ |\mathcal Z|} 2^{nH_Q(Z|Y)}\nonumber\\
& =   &  2^{ -n (R  -
I_Q(X;Z|Y)-\epsilon_n)}\label{eqnarray:appendix 3}
\end{eqnarray}
(\ref{eqn.appendix2}) is true because $Q_{Z|XY}(x^n(i), y^n)\bigcap
\phi^{-1}(i)$, $i=1,2,...,2^{nR_x}$ are disjoint and $\bigcup_i
Q_{Z|XY}(x^n(i), y^n)\subseteq Q_{Z|Y}( y^n)$. Now notice that the
LHS of (\ref{eqn.appendix1}) is at most $2^{n(R_x-R)}\leq 1$, hence
the LHS of (\ref{eqn.appendix1}) is no bigger than $1$. This
together with (\ref{eqnarray:appendix 3}), Lemma~\ref{lemma:Dueck3}
is proved.\hfill$\square$

\vspace{0.1in}

\subsubsection{Proof of~(\ref{eqn.contra2})}\label{sec.appendix.detailforXY} The proof is similar to that of~(\ref{eqn.contra1}).
The difference is that we need the notion of jointlyg good code
books. A code book pair $(c_X, c_Y)$ is good if
\begin{eqnarray}
|c_X\times c_Y \bigcap T^C_\sigma|\leq \frac{|c_X||c_Y|}{4}
\end{eqnarray}
where the joint typical set $T_\sigma=\{(x^n, y^n): D((x^n,
y^n)\|P_X\times P_Y)< \sigma\}$. The rest of the proof are similar
to that in the proof for (\ref{eqn.contra1}). We conclude that
\begin{eqnarray}
\Pr((\widehat m_x,\widehat m_y)= (m_x, m_y) )  &\leq & \frac{1}{4}+
2^{-n(E-\epsilon_n)}+4\sigma\label{eqn.appendix_XYFINAL}
\end{eqnarray}
$\mbox{where } E=\min\limits_{Q_{XYZ}: D(Q_{XY}\|P_X\times P_Y)<
\sigma} D(Q_{Z|XY}\|W_{Z|XY}|Q_{XY}) +|R-I_Q(X,Y;Z)|^+ >0 \mbox {,
for } R_x+R_y> I(X,Y;Z)$.

Again, we need to use a  modified version of Lemma 5 and 3
from~\cite{Dueck_RC} to prove~(\ref{eqn.appendix_XYFINAL}). The
proof is extremely similar to those in Lemma~\ref{lemma:Dueck3}
and~\ref{lemma:Dueck5}. We omit the details here. \hfill $\square$

\end{document}